\documentclass[preprint,showpacs,prb,superscriptaddress]{revtex4}
\usepackage{graphicx}

\newcommand\bea{\begin{eqnarray}}
\newcommand\eea{\end{eqnarray}}
\newcommand\beq{\begin{equation}}
\newcommand\eeq{\end{equation}}

\newcommand{\bib}{\bibitem}

\def\nn{\nonumber}
\def\dg{\dagger}
\def\f{\frac}
\def\la{\langle}
\def\ra{\rangle}
\def\up{\uparrow}
\def\dn{\downarrow}

\def\a{\alpha}
\def\d{\delta}
\def\e{\epsilon}
\def\g{\gamma}
\def\bg{{\bar{\g}}}
\def\G{\Gamma}
\def\bG{\bar{\Gamma}}
\def\l{\lambda}
\def\o{\omega}
\def\r{\rho}
\def\s{\sigma}
\def\S{\Sigma}

\def\p{\partial} 
\def\CH{{\mathcal{H}}}
\def\CG{{\mathcal{G}}}

\def\n{\eta }
\def\tn{\tilde{\eta} }

\def\tc{\tilde{c}}
\def\bG{\bar{G}}

\begin{document}

\title{Non-equilibrium Green's function formalism and the problem of bound 
states}
\author{Abhishek Dhar}
\email{dabhi@rri.res.in}
\affiliation{Raman Research Institute, Bangalore 560080, India}
\author{Diptiman Sen}
\email{diptiman@cts.iisc.ernet.in}
\affiliation{Centre for High Energy Physics, Indian Institute of Science,
Bangalore 560012, India}

\begin{abstract}
The non-equilibrium Green's function formalism for infinitely extended 
reservoirs coupled to a finite system can be derived by solving the equations 
of motion for a tight-binding Hamiltonian. While this approach gives the 
correct density for the continuum states, we find that it does not lead, in 
the absence of any additional mechanisms for equilibration, to a unique 
expression for the density matrix of any bound states which may be present. 
Introducing some auxiliary reservoirs which are very weakly coupled to the 
system leads to a density matrix which is unique in the equilibrium situation,
but which depends on the details of the auxiliary reservoirs in the 
non-equilibrium case.
\end{abstract}
\vskip .5 true cm

\maketitle

\section{Introduction}
\label{sec:intro}

Electronic transport in mesoscopic systems has been studied intensively for 
several years \cite{datta,imry,landauer}. For one-dimensional systems, the 
Landauer-B\"uttiker formalism has played a pivotal role in this subject 
\cite{landauer}. For a wire in which only one channel is available to the 
electrons and the transport is ballistic (i.e., there are no impurities 
inside the wire, and there is no scattering from phonons or from the contacts 
between the wire and the reservoirs at its two ends), the zero-temperature 
conductance is given by $C_0 = 2e^2 /h$ for infinitesimal bias. If there are 
impurities inside the wire which scatter the electrons, then the conductance 
is reduced from $C_0$.

A powerful calculational method for studying electronic transport is the
non-equilibrium Green's function (NEGF) formalism \cite{datta}. The advantage 
of this method is that it treats the infinitely extended reservoirs (leads) 
in an exact way. The derivation of this formalism has been based on the 
Keldysh techniques \cite{keldysh,caroli,datta2,meir1,kamenev,tsukada}. 
Recently a simple derivation of the NEGF results based on a direct solution 
of the equations of motion for a non-interacting system of electrons was 
given in Ref.~\onlinecite{dhar}. This method, based on writing quantum
Langevin equations, was first applied in the case of oscillator
systems by Ford,~Kac and Mazur \cite{fkm}. It has recently been
applied in the context of transport \cite{connell,hanggi,zurch,segal,ingold}.

In the present paper we point out a particular problem that arises while using
the NEGF formalism in a situation where there are bound states and there are 
no additional mechanisms for equilibration (such as electron-phonon 
scattering). We define bound states as states whose wave functions decay 
exponentially as one goes deep into any of the reservoirs. Their energy 
levels lie outside the energy band of all the reservoirs. One expects that 
the NEGF results should reduce to the usual equilibrium results if all the 
reservoirs are kept at the same chemical potential and temperature. This is 
easily shown to be true in the absence of bound states. In the presence 
of bound states, we show that while the contribution of the bound states to 
the equilibrium density matrix can be obtained within NEGF, the procedure is 
subtle and somewhat {\emph {ad hoc}}. It is not clear in this formalism
what the mechanism for equilibration of the bound states is. Moreover, if 
bound states are present, the density matrix is {\emph {not}} unique in the 
non-equilibrium case. Here we show that the equation of motion
approach can be used to obtain a clearer understanding of this problem of 
equilibration of bound states and the non-uniqueness of the non-equilibrium 
steady state. The central results of this paper are as follows. 

(1) We give a simple and general derivation of the NEGF results by the 
equation of motion method for a system without interactions. This is obtained
in two different ways: (i) from the steady state solution of the equations of 
motion, and (ii) from the general solution involving initial conditions. 
 
(2) We show that, in the presence of bound states, the exact solution of the 
wire plus reservoir equations of motion (without any additional sources of 
equilibration) leads to steady states which depend on the initial conditions 
of the wire. 

(3) We show that introducing additional broad-band auxiliary reservoirs (which
are very weakly coupled to the wire) solves the problem of initial condition 
dependence. We obtain the non-equilibrium steady state properties in the 
limit where the coupling strength of the auxiliary reservoirs goes to zero. 
We find that the equilibrium density matrix is then unique and independent of 
the properties of the auxiliary reservoirs. But the non-equilibrium density 
matrix depends on the details of the auxiliary reservoirs and on the way 
in which their couplings (to the wire) are taken to zero.

Bound states have recently been studied in the context of the NEGF
formalism \cite{wang1,wang2}, but to our knowledge, this particular
problem of equilibration has not been addressed earlier. In this paper, we 
deal with electronic transport in non-interacting systems modeled by 
tight-binding Hamiltonians. For simplicity, we only consider spinless 
fermions here although it is quite straightforward to include spin. 

The paper is organized as follows. In Sec.~\ref{sec:negf}, we discuss the 
NEGF formalism and present the expressions for the density matrix in the 
wire and the current. In Sec.~\ref{sec:negfD}, we present a derivation of 
the NEGF results using the equation of motion method. This derivation is 
similar to that of Ref.~\onlinecite{dhar} but is a simplified and more 
generalized version. Starting from the full Heisenberg equations of motion of 
the wire and reservoirs, we derive effective quantum Langevin equations for 
the wire. These equations are solved by Fourier transforms to give the steady 
state solution which leads to expressions for the density matrix and the 
current which are identical to the results obtained from NEGF. 
In Sec.~\ref{sec:bndprob}, we point out the problem of 
equilibration of bound states. In Sec.~\ref{sec:gensol}, we consider the 
general solution of the equations of motion, as opposed to the steady state 
solution obtained in Sec.~\ref{sec:negfD}. This lets us understand better 
the problem of equilibration in the presence of bound states. The question 
of the approach to the steady state (both in cases with or without bound 
states) can also be addressed in this approach. In Sec.~\ref{sec:reso}, we 
describe our method of resolving the problem of equilibration of bound 
states. Namely, we introduce auxiliary reservoirs which are weakly coupled 
to the wire in such a way that the bound states which were earlier localized 
near the wire now extend infinitely into these new reservoirs, and the
original bound state energy levels now lie within the energy band of the 
auxiliary reservoirs. The steady state properties are obtained in the limit 
in which the coupling of auxiliary reservoirs to the wire is taken to zero. 
In Sec.\ref{sec:num}, we present some numerical results, for a system of a 
wire with a few sites coupled to one-dimensional reservoirs, to illustrate 
some of the analytical results. In Sec.~\ref{sec:inter}, we briefly consider 
systems of interacting electrons, and explain why a proper treatment of bound
states is important for computing the current. In Sec.~\ref{sec:disc}, we 
make some concluding remarks.

\section{Non-equilibrium Green's function formalism}
\label{sec:negf}

In this section, we will briefly discuss the NEGF formalism 
\cite{datta,caroli,datta2,meir1,tsukada,dhar} for a system which consists of 
a wire connected to reservoirs which are maintained at different chemical 
potentials or different temperatures. In the NEGF formalism, both the wire and 
the reservoirs are modeled by microscopic Hamiltonians. We will use a 
tight-binding model of non-interacting electrons which we will now describe. 
We use the following notation. For lattice sites anywhere on the system we 
will use the indices $r,s$; for sites on the wire ($W$) we will use the 
integer indices $i,j,k, \cdots$; for sites on the left reservoir ($L$) 
we use the Greek indices $\a, \nu$; finally, for sites on
the right reservoir ($R$) we use the primed Greek indices $\a', \nu'$.
We consider the following Hamiltonian of the full system:
\bea
{\mathcal{H}} &=& \sum_{rs} ~H_{rs} ~c^{\dg}_r c_s \\
&=& \CH_W ~+~ \CH_L ~+~ \CH_R ~+~ {\mathcal{V}}_{L} ~+~{\mathcal{V}}_R ~, 
\nn \\
{\rm where} \quad \CH_W &=& \sum_{lm} ~H^W_{lm} ~c_l^\dg c_m~, ~~~{\CH}_L ~=~
\sum_{\a \nu} ~H^L_{\a \nu} ~c_\a^\dg c_\nu ~,~~~ {\CH}_R ~=~ \sum_{\a' \nu'}~
H^R_{\a' \nu'} ~c_{\a'}^\dg c_{\nu'} ~, \nn \\
{\mathcal{V}}_{L} &=& \sum_{l \a} ~[~ V^L_{l \a} ~c_l^\dg c_\a ~+~ 
{V^L_{\a l}}^{\dg} ~c_\a^\dg c_l ~] ~, \nn \\
{\mathcal{V}}_{R} &=& \sum_{l \a'} ~[~ V^R_{l \a'} ~c_l^\dg c_{\a'} ~+~ 
{{V^R_{\a' l}}^\dg} ~c_{\a'}^\dg c_l ~] ~, \nn
\eea
where $c_r^\dg,~c_r$ denote creation and destruction
operators satisfying the usual fermionic anticommutation relations. The
parts $\CH_W$, $\CH_L$ and $\CH_R$ denote the Hamiltonians of the
isolated wire, left and right reservoirs respectively, while $\mathcal{V}_L$ 
and $\mathcal{V}_R$ describe the coupling of the left and right reservoirs 
to the wire. The main results of NEGF are expressions for the steady state 
current and density matrix in the non-equilibrium steady state. To state 
these results we need a few definitions which we now make. Let $G^+(\o)$ be 
the full single particle Green's function of the system but defined between 
sites on the wire only. (See App.~\ref{appG} for definitions of the various 
Green's functions that will be used). Thus if the wire has
$N$ sites then $G^+$ is a $N \times N$ matrix.
It can be shown (see App.~\ref{appG}) that $G^+$ is given by
\bea
G^+(\o)=\f{1}{\o -H^W-\S_L^+(\o)-\S_R^+(\o)} ~,
\label{self}
\eea
where $\S^+_{L,R}$ are self-energy terms which basically model the effect of
the infinite reservoirs on the isolated wire Hamiltonian. (We will 
work in units in which Planck's constant $\hbar = 1$). The effective wire
Hamiltonian is thus $H^W+\S^+_L+\S^+_R$ which in general will be shown to be
non-Hermitian. The self energies can be written in terms of the isolated
reservoir Green's functions $g^+_{L,R}(\o)$ and the
coupling matrices $V^{L,R}$ (App.~\ref{appG}). We get
\bea
\S^+_L(\o)=V^L~ g^+_L(\o)~ {V^L}^{\dg} ~,~~~~\S^+_R(\o)=V^R~ g^+_R(\o)~ 
{V^R}^{\dg} ~.
\eea

Finally, let us use the following notation for the imaginary parts of the
self energies from the two reservoirs,
\bea
\G_L(\o)=\f{1}{2\pi i} [~\S^-_L-\S^+_L~]=V^L~ \rho_L~ {V^L}^{\dg} ~,~~~~
\G_R(\o)=\f{1}{2 \pi i} [~\S^-_R - \S^+_R~]=V^R~ \rho_R~ {V^R}^{\dg} ~,
\eea
where $\S^-(\o)=V g^-(\o) V^\dg$, and $\r( \o)=-(1/\pi)Im[g^+(\o)]$ is 
the density matrix of an isolated reservoir.

With these definitions, NEGF gives the following expressions for the
density matrix and current:
\bea
n_{lm} &=& \la~ c_m^\dg(t) c_l(t)~ \ra \nn \\
&=& \int_{-\infty}^\infty d\o ~[~(G^+\G_L G^-)_{lm}
f(\o,\mu_L,T_L) ~+~ (G^+\Gamma_R G^-)_{lm} f(\o,\mu_R,T_R)~] ~, \nn \\
J &=& \sum_{m\a} ~\la ~i~[~ V_{m \a}^* ~c^\dg_\a(t) c_m(t) - V_{m \a} ~
c_m^\dg(t) c_\a(t)~]~ \ra \nn \\&=& 2 \pi \int_{-\infty}^\infty d\o ~
Tr[~\G_L G^+ \G_R G^-~]~ (f(\o,\mu_L,T_L) ~-~ f(\o,\mu_R,T_R)) ~,
\label{negfres}
\eea
where $f(\o,\mu,T)=1/[e^{\beta(\o-\mu)}+1]$ denotes the Fermi function.

\section{Derivation of NEGF by ``equation of motion approach'': Quantum 
Langevin equations and solution by Fourier transforms}
\label{sec:negfD}

In this section we give a derivation of the NEGF results using the equation of
motion approach which was developed in Ref.~\onlinecite{dhar}. The derivation 
we present here is a simplified and generalized version of 
Ref.~\onlinecite{dhar}.
This method basically involves writing the full equations of motion of the 
system of wire and reservoirs. The reservoir degrees are eliminated to give 
effective Langevin equations for the wire alone. Finally the Langevin 
equations, which are linear for the case of a non-interacting system, are 
solved by Fourier transformations to obtain the steady state properties. 

The Heisenberg equations of motion for sites on the wire are:
\bea
\dot{c_l}=-i ~\sum_m ~H^W_{lm}~ c_m ~-~ i ~\sum_{\a} ~V^L_{l \a}~ c_\a ~-~ i~
\sum_{\a'} ~V^R_{l \a'}~ c_{\a'} ~,
\label{Esyseq}
\eea
and for sites on the reservoirs are:
\bea
\dot{c}_{\a} &=& -i ~\sum_{\nu} ~H^L_{\a \nu}~ c_\nu ~-~ i ~\sum_l ~
V^{L^\dg}_{\a l}~ c_l ~, \nn \\
\dot{c}_{\a'} &=& -i ~\sum_{\nu'} ~H^R_{\a' \nu'}~ c_{\nu'} ~-~ i ~\sum_l ~
V^{R^\dg}_{\a'l}~ c_l ~.
\label{Ereseq}
\eea
We solve the reservoir equations by treating them as linear equations with
the term containing $c_l$ giving the inhomogeneous part. Using the Green's
functions
\bea
g^+_L(t)=-i~e^{-itH^L}~\theta(t)=\f{1}{2 \pi} \int_{-\infty}^\infty d\o~ 
g^+_L(\o) e^{-i \o t} ~, \nn \\
g^+_R(t)=-i~e^{-itH^R}~\theta(t)=\f{1}{2 \pi} \int_{-\infty}^\infty d\o~ 
g^+_R(\o) e^{-i \o t} ~,
\eea
we get the following solutions for the reservoir equations of motion 
Eq.~(\ref{Ereseq}) (for $t > t_0$):
\bea
c_\a(t) &=& i~ \sum_{\nu} ~[g^+_L (t-t_0)]_{\a \nu}~ c_\nu(t_0) ~+~ 
\int_{t_0}^t dt'~ \sum_{\nu l} ~[g^+_{L}(t-t')]_{\a \nu}~ V^{L^\dg}_{\nu l}~ 
c_l(t') ~, \label{Eres1sol} \\
c_{\a'}(t) &=& i~ \sum_{\nu'} ~[g^+_R(t-t_0)]_{\a' \nu'}~ c_{\nu'}(t_0) ~+~ 
\int_{t_0}^t dt'~ \sum_{\nu' l} [g^+_R (t-t')]_{\a' \nu'}~ V^{R^\dg}_{\nu' l}~
c_l(t') ~.
\label{Eres2sol}
\eea
Plugging these into the equations of motion for the wire Eq.~(\ref{Esyseq}),
we get:
\bea
\dot{c_l} &=& -i ~\sum_m ~H^W_{lm} ~c_m ~-~ i \n_l^L ~-~ i \int_{t_0}^t dt'~ 
\sum_{\a \nu m} V^L_{l \a}~ [g^+_L (t-t')]_{\a \nu}~ V^{L^\dg}_{\nu m}~ 
c_m(t') \nn \\ 
& & -i ~\n_l^R ~-~ i \int_{t_0}^t dt'~ \sum_{\a' \nu' m} ~V^R_{l \a'}~ 
[g^+_R(t-t')]_{\a' \nu'}~ V^{R^\dg}_{\nu' m}~ c_m(t') ~, 
\label{EeqS} \\
{\rm where} \quad \n_l^L &=& i~ \sum_{\a \nu} ~V^L_{l \a}~ 
[g_L^+ (t-t_0)]_{\a \nu}~ c_\nu (t_0) ~,~~~ \n_l^R ~=~ i~ \sum_{\a' \nu'} ~
V^R_{l \a'}~ [g_R^+ (t-t_0)]_{\a' \nu'}~ c_{\nu'}(t_0) ~. \nn \\
& & 
\eea
We have broken up the reservoir contributions into noise and dissipative 
parts; it is clear that the wire equations now have the structure of quantum 
Langevin equations. The properties of the noise terms $\n^L,~\n^R$ can be 
obtained from the condition that at time $t=t_0$, the two reservoirs are 
isolated and described by grand canonical ensembles at temperatures and 
chemical potentials given by $(T_L, \mu_L)$ and $(T_R, \mu_R)$ respectively. 
Thus we find, for the left reservoir, 
\beq
\la ~{\n_l^L}^{\dg} (t) \n_m^L (t')~ \ra ~=~ \sum_{\a \nu \mu \s} ~
V^{L^*}_{l \a} ~[{g_L^+}^*(t-t_0)]_{\a \nu} ~V^L_{m \mu}~ 
[{g^+_L (t'-t_0)}]_{\mu \s} ~\la~ c^\dg_\nu(t_0) c_\s (t_0) ~\ra ~.
\label{Enn}
\eeq
Let $\psi^L_q(\a)$ and $\l^L_q$ denote the eigenvectors and eigenvalues
of the left reservoir Hamiltonian $H^L$ which satisfy the equation 
\bea
\sum_m ~H^L_{\a \nu}~ \psi^L_q(\nu) ~=~ \l^L_q~ \psi^L_q(\a) ~.
\eea 
The equilibrium correlations are given by:
\bea
\la c^\dg_\nu(t_0) c_\s (t_0) \ra = \sum_q \psi^{L^*}_q(\nu)~
\psi^L_q(\s)~ f(\l^L_q,\mu_L,T_L) ~.
\eea
Using this and the expansion $(g^+_L)_{\nu \s}=-i \theta (t-t_0) \sum_q 
\psi^L_q(\nu) ~\psi_q^{L^*}(\s)~ e^{-i \l^L_q (t-t_0)}$ in Eq.~(\ref{Enn}), 
we get
\bea
\la ~{\n^L}^\dg_l(t) \n_m^L (t')~\ra ~=~ \sum_{\a \nu} ~V^{L^*}_{l \a}~ 
[~\sum_q~\psi^{L^*}_q(\a)\psi^L_q(\nu) ~e^{i \l^L_q (t-t')} ~
f(\l^L_q,\mu_L,T_L)~]~ V^{L^T}_{\nu m} ~,
\label{Enono}
\eea 
with similar results for the noise from the right reservoir.
Now let us take the limits of infinite reservoir sizes and let $t_0
\to -\infty$. On taking the Fourier transforms
\bea
\tc(\o)=\f{1}{2 \pi} \int_{-\infty}^\infty dt~ c(t)~ e^{i \o t}~,~~~~
\tn(\o)=\f{1}{2 \pi} \int_{-\infty}^\infty dt~ \n(t)~ e^{i \o t}~,
\eea
we get from Eq.~(\ref{EeqS}):
\bea
\sum_m ~[\o \d_{lm} - H^W_{lm}]~ \tc_m (\o) &=& \sum_m ~(~ [{\S^+_L(\o)}]_{lm}+
[{\S^+_R(\o)}]_{lm}~)~ \tc_m (\o) +\tn_l^L(\o)+\tn^R_l(\o) ~, \nn \\
& & \\
{\rm{where}}~~~~ \S^+_L&=& V^L g^+_L {V^L}^{\dg} ~,~~~~ \S^+_R=V^R g^+_R 
{V^R}^{\dg} ~. \nn
\eea
The terms $\S^+_L,~\S^+_R$ are the same self energies that appear in the NEGF 
formalism and effectively change the Hamiltonian, $H^W$, of the isolated wire,
to $H^W+\S^+_L(\o)+\S^+_R(\o)$. The noise correlations can be obtained from 
Eq.~(\ref{Enono}) and give:
\bea
\la~ \tn_l^{L^\dg}(\o) \tn_m^L(\o')~ \ra &=& (V^L \rho^L {V^L}^{\dg})_{ml}~
f(\o,\mu_L,T_L)~ \d(\o - \o') \nn \\ &=& \G^L_{ml} (\o)~ f(\o,\mu_L,T_L)~ 
\d(\o-\o')~, \\
{\rm{where}}~~~\rho^L_{\a \nu}(\o)&=&\sum_q \psi^L_\a (q)~\psi_\nu^{L^*}(q)~
\d (\o-\l^L_q) ~, \nn
\eea
and similarly for the right reservoir. This is a
fluctuation-dissipation relation and shows how the noise-noise correlations 
are related to the imaginary part of the self energy. Finally, we get the 
following steady state solutions of the equation of motion:
\bea
c_l(t) &=& \int_{-\infty}^\infty~ d\o~ \tc_l (\o)~ e^{-i \o t} ~, 
\label{sswire} \\ 
{\rm with} \quad \tc_l(\o) &=& \sum_m ~{G^+}_{lm}(\o)~[~\tn^L_m(\o)+\tn^R_m 
(\o)~] ~, \\
{\rm and} \quad G^+(\o) &=& \f{1}{\o-H^W-\S^+_L-\S^+_R} ~. 
\eea
For the reservoir variables, we have from Eqs.~(\ref{Eres1sol},\ref{Eres2sol})
\bea
\sum_\a ~V^L_{l \a}~ c_\a(\o)= \tn^L_l (\o) + \sum_{mn} [{\S^+_L(\o)}]_{l m}~ 
G^+_{m n}(\o)~ [~ \tn^L_n (\o) +\tn^R_n (\o)~] ~, \\
\sum_{\a'} ~V^R_{l \a'}~ c_{\a'}(\o)= \tn^R_l (\o) + \sum_{mn} 
[{\S^+_R (\o)}]_{l m}~ G^+_{m n}(\o)~ [~ \tn^L_n (\o) +\tn^R_n (\o)~] ~.
\label{ssres}
\eea

\subsubsection{Steady state current and densities}

{\bf {Current}}: From a continuity equation, it is easy to see that the
net current in the system is given by the following expectation value,
\bea
J = \sum_{m\a} \la~ i~[~ V_{m \a}^{L^*}~ c^\dg_\a(t)~ c_m(t) - V^L_{m \a}~ 
c_m^\dg(t) ~c_\a(t)~]~ \ra= \sum_{m \a} 2~ Im[~ V^L_{m \a}~ \la~ c_m^\dg (t)~
c_\a(t)~ \ra~ ].
\label{Ecur1}
\eea
{}From the steady state solution (Eqs.~\ref{sswire}-\ref{ssres}), we get
\bea
&& \sum_{m \a} V^L_{m \a}~ \la c^\dg_m(t)~ c_\a(t)\ra = \sum_{m \a} 
V^L_{m \a}~ \int_{-\infty}^\infty d\o ~\int_{\infty}^\infty d\o'~ 
e^{i (\o-\o') t}~ \la ~\tc^\dg_m (\o)~ \tc_\a (\o')~ \ra \nn \\
&& = \sum_m ~ \int_{-\infty}^\infty d\o~ \int_{\infty}^\infty d\o'~ 
e^{i (\o-\o') t}~ \sum_l ~\la ~G^-_{lm} (\o)~ [~\tn_l^{L^\dg} (\o)
+\tn^{R^\dg}_l (\o)~] \nn \\ 
&& ~~~~~~~~~~~~~~~~~~~ \times ~[ \tn_m^L (\o')+ \sum_{jk} ~
[~{\S^+_L(\o')}]_{mj}~ G^+_{jk}(\o')~ (~\tn^L_k(\o')+ \tn^R_k(\o')~)~]~ \ra ~.
\eea
Let us consider that part of the above expression which depends only
on the temperature and chemical potential of the right reservoir. This is given by
\bea
&&\int_{-\infty}^\infty d\o~ \int_{-\infty}^\infty d\o'~ e^{i(\o-\o') t} ~
\sum_{lmjk} ~G^-_{lm} (\o)~[ {\S^+_L(\o')}]_{mj}~ G^+_{jk}(\o') ~\la
~\tn^{R^\dg}_l (\o) ~\tn^R_k(\o'))~ \ra \nn \\
&&= \int_{-\infty}^\infty d\o ~\sum_{lmjk} ~G^-_{lm} (\o) [{\S^+_L(\o)}]_{mj}~
G^+_{jk}(\o)~ {[\Gamma_R (\o)]}_{kl}~ f(\o,\mu_R,T_R) ~ \nn \\
&&= \int_{-\infty}^\infty d \o ~Tr[G^- (\o)~ \S_L^+(\o)~G^+(\o)~\G_R (\o)] ~
f(\o, \mu_R,T_R)~. \nn
\eea
Taking the imaginary part of this and using this in Eq.~(\ref{Ecur1}), we
find the contribution of the right reservoir to the current to be
\bea
J_R=-2 \pi \int_{-\infty}^\infty d\o ~Tr[\Gamma_L G^+ \Gamma_R G^-] 
~f(\o,\mu_R,T_R) ~.
\eea
[In deriving this result we used the identities $G^{+^\dg}=G^-,~\S^{+^\dg}=
\S^-,~\G^\dg=\G$ and the fact that $(Tr[M])^*=Tr[M^\dg]$ for any matrix $M$].
It is clear that on adding the contribution of the left reservoir, we
will get the net current (left-to-right) to be
\bea
J=2 \pi \int_{-\infty}^\infty d\o ~Tr[\Gamma_L G^+ \Gamma_R G^-] 
~(f(\o,\mu_L,T_L)-f(\o,\mu_R,T_R)) ~.\label{jft}
\eea

{\bf {Densities}}: The density matrix is given by the expression
\bea
n_{lm} &=& \la~ c_m^\dg(t)~ c_l(t)~ \ra \nn \\
&=& \int_{-\infty}^\infty d\o \int_{-\infty}^\infty d\o'~ e^{i (\o -\o') t}~
\sum_{kj} ~G_{km}^-(\o)~ \la~ [\tn_k^{L^\dg} (\o)+\tn_k^{R^\dg} (\o)]~
[ \tn^L_j (\o') +\tn^R_j (\o')]~\ra~ G^+_{lj}(\o') \nn \\
&=& \int_{-\infty}^\infty d\o ~[~(G^+\Gamma_L G^-)_{lm}
~f(\o,\mu_L,T_L)+(G^+\Gamma_R G^-)_{lm}~ f(\o,\mu_R,T_R)~] ~.\label{nft} 
\eea
The results in Eq.~(\ref{jft},\ref{nft}) are identical to the results
from NEGF given in Sec.~\ref{sec:negf}.
For $\mu_L=\mu_R$ and $T_L=T_R$, we get the expected equilibrium value
[see App.~(\ref{appeq})] {\emph {provided that there are no bound states}}. 

\section{Problems with bound states}
\label{sec:bndprob}

A desirable property of the NEGF results is that they should reduce to the 
standard equilibrium results for the case when both reservoirs are at the same
chemical potential and temperatures, {\emph {i.e.,}} for $\mu_L=\mu_R= \mu$
and $T_L=T_R=T$. For this case the current vanishes, which is the correct 
result expected from equilibrium. For the density, we get from 
Eq.~(\ref{negfres}):
\bea
n_{lm} &=& \int_{-\infty}^\infty d\o~ [G^+(\o)~ (~\G_L(\o) +\G_R(\o)~)~ 
G^-(\o)]_{lm}~ f(\o,\mu,T) ~, \label{negfeq} \\
{\rm with}~~~~G^+(\o) &=& \f{1}{\o-H^W-\S^+_L(\o)-\S^+_R(\o)} ~. \label{goute}
\eea 
The expected equilibrium result can be obtained using the grand canonical 
ensemble, and we get (see App.~\ref{appeq}):
\bea
n^{eq}_{lm} &=& \int_{-\infty}^\infty d \o~ [G^+ ~( \G_L + \G_R )~
G^-]_{lm}~ f(\o,\mu,T) ~+ ~\int_{-\infty}^\infty d \o ~\f{\e}{\pi}~ 
[G^+ G^-]_{lm}~ f(\o, \mu,T) ~, ~~~~~~~ \label{eqdens} \\
&&{\rm where} ~~G^+(\o) = \f{1}{\o+ i \e-H^W-\S^+_L(\o)-\S^+_R(\o)} ~.
\eea
Note that here we retain the $i\e$ factor in the Green's function. The $i\e$
factor in Eq.~(\ref{goute}) can be dropped since, for $\o$ in the range of 
interest in Eq.~(\ref{negfeq}), the self energies $\S^+_L$ and $\S^+_R$ have 
finite imaginary parts. (Whenever 
we introduce $\e$ in an equation, it is understood that it is an infinitesimal
positive quantity which has to be taken to zero at the end of the calculation).
The first piece in Eq. (\ref{eqdens}) is identical to the NEGF prediction, 
while it can be shown that the second piece vanishes {\emph {only}} if there 
are no bound states. Thus in the absence of bound states we verify that the 
NEGF results reduce to the correct equilibrium results. 

Let us discuss now the case when there are bound states. Here we refer to 
bound states of the full system of wire and reservoirs. The bound states have 
energies lying outside the range of the reservoir levels, and the wave 
functions corresponding to them decay exponentially as one goes deep into
the reservoirs. They are obtained as real solutions of the equation
\bea
[H^W+\S^+_L(\l_b)+\S^+_R(\l_b)] ~\Psi_b ~=~ \l_b \Psi_b ~.
\eea 
{}From the form of the Green's function, it is clear that the second term in
Eq.~(\ref{eqdens}) is non-vanishing whenever the above equation has real
$\l_b$ solutions, and it can be shown to reduce to the form 
\bea
n^{eq-bnd}_{lm} ~=~ \sum_b ~\Psi_b(l) \Psi^*_b(m) ~f(\l_b,\mu,T)~. 
\label{eqbnd}
\eea 
Thus there seems to be a problem of equilibration in the NEGF formalism 
whenever bound states are present. This problem can be fixed in the following 
way. In the NEGF results let us add an extra infinitesimal part $\e_a/\pi$ to 
the matrix $\G_a(\o)$ corresponding to the $a^{\rm th}$ reservoir ($a=L,~R$). 
Thus the NEGF result for the density matrix is modified to
\bea
n_{lm} ~=~ \la c_m^\dg(t) c_l(t) \ra 
&=& \int_{-\infty}^\infty d\o~ [~(G^+\G_L G^-)_{lm}~ 
f(\o,\mu_L,T_L)+(G^+\Gamma_R G^-)_{lm}~ f(\o,\mu_R,T_R)~] \nn \\ &+& 
\int_{-\infty}^\infty d\o~ [~\f{ \e_L}{\pi}~ (G^+ G^-)_{lm}~ f(\o,\mu_L,T_L) + 
\f{\e_R}{\pi} ~(G^+ G^-)_{lm} ~f(\o, \mu_R, T_R)~], \nn \\
{\rm with} ~~~G^+(\o)&=& \f{1}{\o-H^W-\S^+_L(\o)-\S^+_R(\o) + i 
(\e_L+\e_R)} ~.
\eea 
The second integral gets contributions from bound states. Using the identity
\bea
\lim_{{\e_{L,R}} \to 0} ~\f{\e_a}{(\o-\l_b)^2 + (\e_L+\e_R)^2} ~=~
\lim_{{\e_{L,R}} \to 0} ~\pi \f{\e_a}{\e_L+\e_R} \d( \o-\l_b) ~,
\eea
we finally get the following contribution from the bound states,
\bea
n^{bnd}_{lm}=\sum_b ~\lim_{{\e_{L,R}}\to 0}~ \f{\e_L f(\l_b,\mu_L,T_L)+
\e_R f(\l_b,\mu_R,T_R)}{\e_L+\e_R} ~\Psi_b(l) \Psi_b^*(m) ~.
\label{negfbnd}
\eea
For $\mu_L=\mu_R=\mu$ and $T_L=T_R=T$, we now get the expected equilibrium 
result of Eq.~(\ref{eqbnd}). For the non-equilibrium case, however, the 
situation is somewhat unsatisfactory since the density matrix depends on the 
ratio $\e_L/\e_R$, and so the bound state contribution is ambiguous. More 
importantly, in this approach {\emph {it is not at all clear as to what the 
exact physical mechanism for equilibration of the bound states is}}.

In the following sections, we will examine this particular question of
equilibration of bound states more carefully.

\section{General solution of equations of motion}
\label{sec:gensol}

In this section, we will consider the general solution of the
equations of motion. Unlike the Fourier transform solution obtained in
Sec.~\ref{sec:negfD}, the general solution also involves the {\emph {initial
conditions of the wire}}. The advantage of the equations of motion method 
is that it can address issues such as that of approach to the steady state,
it can be numerically implemented (after truncating the reservoirs to a
finite number of sites), and it does not rely on any {\emph {ad hoc}} $i\e$
prescriptions for the reservoirs. We will show that the expression
for the density matrix can be derived without any difficulties
if there are no bound states. Further, the problems which arise if
there is a bound state become quite clear in this approach.

Let us again consider a system in which a wire is connected to two reservoirs,
with
\beq
\CH ~=~ \sum_{r,s} ~H_{rs}~c_r^\dg ~c_s ~.
\eeq
As before we will use the label $r,s$ to denote any site in the system, and 
$j$, $\a$ and $\a'$ to denote sites in the wire and left and right reservoirs 
respectively. Instead of eliminating the reservoir degrees of motion and 
writing Langevin equations for the wire, we will now deal with the full
Heisenberg equations of motion for the system which is given by
\bea
\dot{c}_r ~=~ -i ~\sum_s ~H_{rs} ~c_s~.
\eea

At the initial time $t=0$, we assume that the two reservoirs are in thermal
equilibrium at temperatures and chemical potentials $T_{L,R}$ and 
$\mu_{L,R}$, and the couplings $V^{L,R}$ are zero.
As before, let $\l^a_q$ and $\psi^a_q$, with $a=L,R$, denote
the eigenvalues and eigenfunctions of the isolated reservoir Hamiltonians.
At $t=0$, the density matrix for sites on the left and right
reservoirs are given respectively by
\bea
\la c^\dg_\a(0) c_\nu (0) \ra &=& \sum_q ~\psi^{L^*}_q(\a) \psi^L_q(\nu) ~
f(\l^L_q,\mu_L,T_L) ~, \nn \\
\la c^\dg_{\a'}(0) c_{\nu'} (0) \ra &=& \sum_q ~\psi^{R^*}_q(\a') 
\psi^R_q(\nu') ~f(\l^R_q,\mu_R,T_R) ~.
\label{initden}
\eea
We also assume that at $t=0$, there are no correlations between the two
reservoirs and between the wire and the reservoirs. Thus
\bea
\la ~c^\dg_l (0)~ c_\a (0) ~\ra ~=~ \la ~c^\dg_l (0) ~c_{\a'} (0)~\ra ~=~ 
\la ~c^\dg_\a (0)~c_{\a'} (0)~\ra ~=~ 0~.
\eea
Finally, for 
sites in the wire, we cannot unambiguously assign a value to the density 
matrix $\la c_l^\dg (0) c_{m} (0) \ra$ at $t=0$, since the wire is isolated
from everything else at that time. Under certain conditions, we will see that
in the steady state (defined as the limit $t \to \infty$), the density matrix 
of the wire will turn out to be independent of its value at $t=0$.

Let us now suddenly switch on the couplings $V_a$ of the wire to the
reservoirs at $t=0$. For $t>0$, the solution of the equations of motion is
given in matrix notation by
\beq
c (t) ~=~ i~ \CG^+(t) ~c(0) ~,
\eeq
where $\CG^+(t) = -ie^{-i H t} \theta(t)$, and $c(t)=[c_1,~c_2...,~
c_{N_s}]^T$, where $N_s$ is the total number of sites in entire system. 
For a point $l$ in the wire, we then have
\bea
c_l (t) ~=~ \sum_m ~\CG^+_{lm}(t)~ c_{m} (0) ~+~ \sum_\nu ~\CG^+_{l \nu}~ 
c_{\nu} (0) ~+~ \sum_{\nu'} ~\CG^+_{l \nu'}~ c_{\nu'} (0) ~.
\eea
Hence the density matrix in the wire is given by
\bea
n_{lm} (t) & = & \la c^\dg_{m} (t) c_l (t) \ra \nn \\
&=& ~\sum_{ij} ~\CG^+_{lj} (t) ~\la c^\dg_i (0) c_j (0) \ra ~\CG^-_{im} (t) ~
+~ \sum_{\a \nu} ~\CG^+_{l \a} (t)~ \la c^\dg_{\nu}(0) c_{\a}(0) \ra ~
\CG^-_{\nu m} (t) \nn \\
&& ~+~ \sum_{\a' \nu'} ~\CG^+_{l \a'} (t)~ \la c^\dg_{\nu'}(0) c_{\a'}(0) 
\ra~ \CG^-_{\nu' m} (t) ~.
\label{nlm}
\eea
Now let $\Psi_Q(r)$ and $\l_Q$ denote the eigenfunctions and eigenvalues
respectively of the system of coupled wire and reservoirs. Thus
\bea
\sum_s ~H_{rs} ~\Psi_Q(s) ~=~ \l_Q \Psi_Q(r) ~.
\eea
Then the full Green's function can be written in the following way (for 
$t >0$):
\bea
\CG^+_{rs}(t) &=& -i ~\sum_Q ~\Psi_Q(r) \Psi^*_Q(s) ~e^{-i \l_Q t} \nn \\
&=& -i ~\sum_b ~\Psi_b(r) \Psi^*_b(s) ~e^{-i \l_b t} ~-~ i ~\int d \o ~
\r^c_{rs} (\o) ~e^{-i \o t} ~,
\label{large}
\eea
where $\Psi_b,~\l_b$ refer to bound states, and the density matrix $\r^c$ is
given by a sum over the extended (continuum) states of the system
$\r^c_{rs}(\o) = \sum_Q^c \Psi_Q(r) \Psi_Q^*(s) \d (\o-\l_Q)$. In the limit
$t \to \infty$, the second term vanishes (this follows from the
Riemann-Lebesgue lemma, see Ref.~\onlinecite{bender}), and we get a
contribution only from bound states. Thus
\bea
\lim_{t \to \infty} \CG^+_{rs}(t) ~=~ -i ~\sum_b ~\Psi_b(r) \Psi^*_b(s)~
e^{-i \l_b t} ~.
\label{Glt}
\eea
Thus if there are no bound states in the fully coupled system then, for any
two sites of the system, $\CG^+_{rs} (t)$ vanishes as $t \to \infty$. From
this it follows that the contribution of any {\it finite} number of terms
in Eq.~(\ref{nlm}) vanishes in the steady state. Since the wire consists of
a finite number of sites, this means that the initial density matrix of the
wire $\la c^\dg_i(0) c_j(0) \ra $ will have no effect on the steady state
density matrix $n_{lm} (t \to \infty )$. The reason that
Eq.~(\ref{nlm}) does not vanish in the steady state is that it gets
contributions from the reservoirs which have an infinite number of sites.
On the other hand the situation is quite different if there are one
or more bound states in the problem. In that case, clearly, individual matrix
elements of $\CG^+$ may not vanish in the steady state, and the initial
state of the wire makes a contribution to the long time density matrix.

We now look at the contribution of the reservoirs in Eq.~(\ref{nlm}). Dropping
the subscript $a$ for the moment, let us look at the contribution of any one
of the reservoirs. Using Eq. (\ref{initden}), this can be written as
\bea
n^{res}_{lm} (t) &=& \sum_{\a \nu} ~\int_{-\infty}^\infty ~{d\o}~
f (\o, \mu , T) ~\CG^+_{l \a} (t) ~\sum_q ~\psi_q (\a) \psi^*_q (\nu)~
\d (\o-\l_q ) ~\CG^-_{\nu m} (t) \nn \\
& \equiv & \int_{-\infty}^\infty {d\o} ~ f (\o, \mu, T) ~ I_{lm} (\o,t) ~,
\label{najj} \\
{\rm where}~~~I_{lm}(\o,t) &=& \sum_{\a \nu} \CG^+_{l \a} (t)~ [~\sum_q ~
\psi_q (\a) \psi^*_q (\nu) ~\d (\o-\l_q )~]~ \CG^-_{\nu m} (t) ~. \nn
\eea
We now transform to the frequency dependent Green's function using the
relation $\CG^+(t) ~ =~ \int_{-\infty}^\infty \f{d\o}{2\pi} \CG^+(\o)
e^{-i \o t}$. As shown in App.~\ref{appG}, we can express the Green's
function elements $\CG^+_{j\nu}(\o)$ in terms of $G^+_{ij}(\o)$ and the 
isolated reservoir Green's functions $g^+_{\a \nu}(\o)$. We get
\bea
\CG^+_{l \a } (\o) &=& \sum_{j \nu} ~G^+_{lj} (\o)~ V_{j \nu}~ g^+_{\nu \a} 
(\o) ~, \nn \\
{\rm and} \quad \CG^-_{\a l} (\o) &=& \sum_{j \nu} ~g^-_{\a \nu} (\o)~ 
V^\dg_{\nu j} ~G^-_{jl} (\o) ~.
\label{res1}
\eea
We will also use the following result which follows from an
eigenfunction expansion of $g^+ (\o)$,
\bea
\sum_{\a } ~g^+_{\nu \a}(\o) \psi_q (\a) ~=~ \sum_\a ~[\sum_{q'}
\f{\psi_{q'}(\nu) \psi^*_{q'}(\a)}{\o+i \e - \l_{q'}}] ~\psi_{q}(\a) ~=~
\f{\psi_{q}(\nu)}{\o+i \e - \l_{q}} ~.
\label{res2}
\eea
Using Eqs.~(\ref{res1}-\ref{res2}), we can write $I_{lm} (\o,t)$ in 
Eq.~(\ref{najj}) as
\bea
&& I_{lm}(\o,t) \nn \\
&=& \f{1}{2 \pi} \int d \o' ~\sum_{j \nu \a i} ~\f{G^+_{lj}(\o') 
e^{-i\o't}}{\o'+i \e - \o} V_{j \nu}~[~ \sum_q \psi_q(\nu) \psi^*_q(\a) 
\d (\o- \l_q)~]~ V^\dg_{\a i} ~\f{1}{2 \pi} \int d \o'' \f{G^-_{im}(\o'') 
e^{i\o''t}}{\o''-i \e - \o} \nn \\
&=& \f{1}{2 \pi} \int d \o' ~\sum_{ji} ~\f{G^+_{lj}(\o') e^{-i\o't}}{\o'
+i \e - \o}~ [V \r(\o) V^\dg]_{ji} ~\f{1}{2 \pi} \int d \o'' ~ 
\f{G^-_{im}(\o'') e^{i\o''t}}{\o''-i \e - \o} ~.
\label{ijj}
\eea
The first integral in the above equation can be evaluated as follows:
\bea
\f{1}{2 \pi} \int d \o' \f{G^+_{lj}(\o') e^{-i\o't}}{\o'+i \e - \o} &=& 
\sum_Q ~\Psi_Q(l) \Psi_Q^*(j) ~\f{1}{2 \pi}\int_{-\infty}^\infty d \o' ~
\f{e^{- i \o' t}}{(\o'+i \e-\o) (\o'+i \e'-\l_Q)} \nn \\
&=& -iG^+_{lj}(\o)e^{-i \o t} ~+~ i \sum_Q ~\Psi_Q(l) \Psi_Q^*(j)~ 
\f{e^{-i \l_Q t}}{\o+i \e -\l_Q} ~.
\eea
In the limit $t \to \infty$, only the bound states contribute to the
summation (over $Q$) in the expression above. Hence we get
\bea
\lim_{t \to \infty}~ \f{1}{2 \pi} \int d \o'~ \f{G^+_{lj}(\o') e^{-i\o't}}{\o'
+i \e - \o} = -i~G^+_{lj}(\o)~e^{-i \o t} +i~ \sum_b ~\Psi_b(l) \Psi_b^*(j)~
\f{e^{-i \l_b t}}{\o-\l_b}~. 
\label{gomit}
\eea
Note that we have dropped the $i \e$ factor in the denominator of the second
term. This is because of the following reason. In the expression 
Eq.~(\ref{ijj}), the presence of the density of states of the
reservoirs $\r(\o)$ means that we will be interested only in values of 
$\o$ lying in the continuum of reservoir eigenvalues. Using Eq.~(\ref{gomit}),
we finally get the following result for the contribution from
the $a^{\rm{th}}$ reservoir to the density matrix in the long time limit:
\bea
n^a_{lm}&=& \int_{-\infty}^\infty d \o ~[~G^+(\o) \Gamma^a (\o)
G^-(\o)~]_{lm}~ f(\o,\mu_a,T_a) \nn \\
&& +~ \sum_{bb'}e^{-i(\l_b-\l_{b'})t} ~\sum_{ji} ~\Psi_b(l) \Psi^*_b(j) 
\Psi_{b'}(i) \Psi_{b'}^*(m) ~\int_{-\infty}^{\infty} d \o~ 
\f{[~\Gamma_a(\o)~]_{ji}~ f(\o,\mu_a,T_a)}{(\o-\l_b) (\o-\l_{b'})} ~, \nn \\
&&
\label{rescont}
\eea
where $b,b'$ denote bound states, and we have again dropped terms in which
the integrands contain time-dependent oscillatory factors.

In the absence of bound states, we see that the contribution of the
reservoirs is identical to that given by NEGF and also
by the Fourier-transform solution in Sec.~\ref{sec:negfD}.
However, in the presence of bound states, there is an extra
contribution which is not obtained in the other methods.
Importantly, this extra part is, in general, time-dependent so that the
system never reaches a stationary state. Also, we do not recover the
equilibrium results for the case when all reservoirs are kept at equal
temperatures and chemical potentials. The reason for this is that the
reservoirs are not able to equilibrate bound state energy levels since
these levels lie outside the range of the reservoir band of energies.

\section{A solution to the bound state problem}
\label{sec:reso}

One way to solve the bound state equilibration problem is to introduce two
auxiliary reservoirs $\bar L$ and $\bar R$; these must have the same
temperature and chemical potentials as the original reservoirs $L$ and $R$,
but they must have a large enough bandwidth so that any bound states of the
original Hamiltonian $\CH$ lie within that bandwidth. (Briefly, the idea
is that since the bound states of $\CH$ are no longer bound states in the 
expanded system, the arguments given in Sec.~\ref{sec:bndprob} imply that 
the density matrix of the wire will no longer depend
on the initial density matrix). The two auxiliary reservoirs must be coupled
very weakly to the wire, so that they do not greatly alter the energies of
$\CH$ and the corresponding wave functions within the wire and the original
reservoirs. We will eventually take the limit of that coupling going to zero.
The main purpose of the auxiliary reservoirs is to equilibrate
any bound states which $\CH$ might have. Let us now see how all this works.

We consider a new Hamiltonian of the form
\bea
\CH_{\rm new} &=& \CH ~+~ \CH_{\bar L} ~+~ \CH_{\bar R} ~+~
{\mathcal V}_{\bar L} ~+~ {\mathcal V}_{\bar R} ~, \\
{\rm where} \quad {\mathcal{H}} &=&\CH_W+\CH_L+\CH_R+{\mathcal{V}}_{L}+
{\mathcal{V}}_{R} ~. \nn
\eea
$\CH_{\bar a}$ denotes the Hamiltonian of the auxiliary reservoirs
(${\bar a} = {\bar L}, {\bar R}$), and ${\mathcal V}_{\bar a}$ denotes the
coupling of those reservoirs to the wire. The auxiliary reservoirs will also
be taken to be lattice systems with tight-binding Hamiltonians,
but with a hopping amplitude which is sufficiently large.
We will take the auxiliary reservoir $\bar L$ to have
the same temperature and chemical potential $T_L$ and $\mu_L$ as the
reservoir $L$, and similarly for the reservoirs $\bar R$ and $R$.

Following the derivation of Sec.~\ref{sec:negfD}, we find that the full Green's
function on the wire is now given by 
\beq
{\bar{G}}^+ (\o) ~=~ \f{1}{\o ~-~ H^W ~-~ \S^+_L(\o) ~-~ \S^+_R(\o) ~-~ 
\S^+_{\bar L}(\o) ~-~ \S^+_{\bar R}(\o)} ~.
\eeq
With the auxiliary reservoirs present, there are no longer any bound
states, and the density matrix in the wire can be written as
\bea
n_{lm} = & & \int_{-\infty}^\infty d\o ~[\bG^+ (\G_L+\G_{\bar L}) \bG^-]_{lm}~
f(\o, \mu_L ,T_L) \nn \\
& & +~ \int_{-\infty}^\infty d\o ~[\bG^+ (\G_R+\G_{\bar R}) \bG^-]_{lm} 
f(\o, \mu_R ,T_R) ~. \nn 
\eea
We want to eventually take the limit in which the couplings of the wire to the
auxiliary reservoirs $V_{\bar a}$ go to zero. We will therefore treat these 
couplings perturbatively. Let us first break up the above integral over $\o$ 
into two parts, one with $\o$ going over the range of the original
reservoir band, and the other containing the remaining part over the
range of the auxiliary reservoir. We assume that the original
reservoir bandwidths are in the range $[E_1,E_2]$
while that of the auxiliary reservoirs are $[{\bar{E}}_1, {\bar{E}}_2]$;
we choose the latter such that it contains the original bandwidth,
{\emph {i.e.}}, ${\bar{E}}_1< E_1$ and ${\bar{E}}_2 > E_2$. Thus we write:
\bea
n_{lm}&=& n^1_{lm} +n^2_{lm} ~, \nn \\
n^1_{lm} &=& \int_{E_1}^{E_2} d\o ~[\bG^+ (\G_L+\G_{\bar L}) \bG^-]_{lm}
f(\o, \mu_L ,T_L) + \int_{E_1}^{E_2} d\o ~[\bG^+ (\G_R+\G_{\bar R}) \bG^-]_{lm}
f(\o, \mu_R ,T_R) , \nn \\
n^2_{lm} &=&\int_{\bar{E}_1}^{E_1} d\o ~[\bG^+ \G_{\bar L} \bG^-]_{lm}
f(\o, \mu_L ,T_L) ~+~ \int_{\bar{E}_1}^{E_1} d\o ~[\bG^+ \G_{\bar R} 
\bG^-]_{lm} f(\o, \mu_R ,T_R) \nn \\
&& +~ \int_{E_2}^{\bar{E}_2} d\o ~[\bG^+ \G_{\bar L}
\bG^-]_{lm} f(\o, \mu_L ,T_L) ~+~ \int_{E_2}^{\bar{E}_2} d\o ~[\bG^+
\G_{\bar{ R}} \bG^-]_{lm} f(\o, \mu_R ,T_R) ~.
\eea
Now, in the first part ($n^1$) we note that both $\S^+_L$ and $\S^+_R$
have finite imaginary parts. Hence on taking the limit $V_{{\bar L},{\bar R}}
\to 0$, we immediately get $\G_{{\bar L},{\bar R}} \to 0$ and $\bG^+(\o)=
G^+(\o)$. Hence
\bea
n^1_{lm} ~=~ \int_{E_1}^{E_2} d\o ~[G^+ \G_L G^-]_{lm} ~f(\o, \mu_L ,T_L) ~+~ 
\int_{E_1}^{E_2} d\o ~[G^+ \G_R G^-]_{lm} ~f(\o, \mu_R ,T_R) ~,
\eea
which is just the usual reservoir contribution.

We will now show that the second part contributes to bound states only. As 
before, let us denote the eigenvalues and eigenfunctions of the combined system
of the wire and reservoirs (but not the auxiliary reservoirs) by $\l_Q$
and $\Psi_Q$, where the label $Q$ can have both a continuous
and a discrete part (corresponding to bound states). We will assume
that the auxiliary reservoirs have been chosen such that all the
values of $\l_Q$ lie within the bandwidth of the auxiliary reservoirs.
The coupling to the auxiliary reservoirs cause the energy levels to shift 
from $\l_Q$ to $\l_Q + \d \l_Q$. To first order in $V_{\bar a}$, we find that
\beq
\d \l_Q ~=~ \la \Psi_Q | \S^+_{\bar L} (\l_Q) + \S^+_{\bar R} (\l_Q) |
\Psi_Q \ra ~=~ \sum_{jk} ~\Psi^*_Q (j) ~[\S^+_{\bar L} +\S^+_{\bar R}]_{jk}~
\Psi_Q (k)~.
\eeq
The imaginary part of $\d \l_Q$ is therefore given by
\beq
{\rm Im} ~\d \l_Q= - {\pi} \sum_{j, k} ~\Psi^*_Q (j) ~[\G_{\bar L} + 
\G_{\bar R}]_{j, k}~ \Psi_Q (k) ~.
\label{imde}
\eeq
Thus for the auxiliary reservoir $\bar a$ we get
\bea
\sum_{km} ~\bG^+_{j k} [\G_{{\bar a}}]_{k m} \bG^-_{m l} ~=~ \sum_{km}~\sum_Q~
\f{\Psi_Q (j) \Psi^*_Q (k)}{\o - \l_Q - \d \l_Q }~ [\G_{\bar a}]_{ k m}~ 
\sum_{Q'}~\f{\Psi_{Q'} (m) \Psi^*_{Q'} (l)}{\o -\l_{Q'} -\d \l^*_{Q'}} ~.
\label{nbara}
\eea
Since we are considering values of $\o$ outside the range $[E_1,E_2]$,
it is clear that, in the limit $\G_{\bar{a}} \to 0 $, only bound
states with $\l_Q=\l_{Q'}=\l_b$ will contribute. The above then gives
\bea
\lim_{\G_{\bar{a}}\to 0 } ~\sum_{km} ~\bG^+_{j k} [\G_{{\bar a}}]_{k m}
\bG^-_{m l}
&=& \sum_{km} ~\sum_b~\f{\Psi_b (j) \Psi^*_b (k) ~[\G_{{\bar a}}]_{km} ~
\Psi_b (m) \Psi^*_b (l)}{(\o -\l_b - ~{\rm Re} ~\d \l_b)^2 ~+~ ({\rm Im} ~
\d \l_b)^2} \nn \\
&=& \sum_b~\f{\la \Psi_b | \G_{\bar a} | \Psi_b \ra}{\la \Psi_b | \G_{\bar L}
+ \G_{\bar R} | \Psi_b \ra} ~\d (\o - \l_b) ~\Psi_b (j) ~\Psi^*_b (l) ~,
\label{small}
\eea
where we have used the fact that both real and imaginary parts of $\d
\l_b$ are small, with the imaginary part given by Eq.~(\ref{imde}).

Putting everything together, we finally find that the total density matrix 
in the wire due to all four reservoirs is given by
\bea
n_{lm} &=& ~\int_{E_1}^{E_2} d\o~ [~ (~G^+ (\o) \G_L (\o) G^-(\o)~)_{lm} ~
f_L (\o) + (~G^+ (\o) \G_R (\o) G^- (\o)~ )_{lm} ~f_R (\o)~ ] \nn \\
& & +~ \sum_b ~ \f{\la \Psi_b | \G_{\bar L} | \Psi_b \ra f_L (\l_b) ~+~ \la
\Psi_b | \G_{\bar R} | \Psi_b \ra f_R (\l_b)}{\la \Psi_b | \G_{\bar L} ~+~
\G_{\bar R} | \Psi_b \ra} ~\Psi_b (l) \Psi^*_b (m) ~,
\label{final}
\eea
where the sum over $b$ runs over all the bound states of $H$. In the
equilibrium case [i.e., for $f_L (\l_b) = f_R (\l_b)$] Eq.~(\ref{final})
shows that the contribution of a bound state $b$ to the density matrix is
independent of details of the auxiliary reservoirs such as the quantities
$\la \Psi_b | \G_{\bar L} | \Psi_b \ra$ and $\la \Psi_b | \G_{\bar R} | \Psi_b
\ra$. But in the 
non-equilibrium case [i.e., for $f_L (\l_b) \ne f_R (\l_b)$], the contribution
of a bound state {\emph{does depend on details of the auxiliary reservoirs}},
namely, on the ratio $\la \Psi_b | \G_{\bar L} | \Psi_b \ra / \la \Psi_b |
\G_{\bar R} | \Psi_b \ra$. This is perhaps not surprising. Some quantities
in equilibrium statistical mechanics may be independent of the mechanism for
equilibration, while the same quantities in non-equilibrium statistical
mechanics may depend on the details of that mechanism.

The similarity between Eq.~(\ref{negfbnd}) and the bound state contribution in
Eq.~(\ref{final}) implies that we can think of the auxiliary reservoirs as
providing a justification for the $i\e_a$ prescription which was introduced
in the NEGF formalism in Sec.~\ref{sec:negf}.

We note that bound states do not carry current, and so the expressions for
current remain unchanged. Later we will point out that in the presence of
electron-electron interactions, even the current is likely to be affected
by bound states.

\section{Numerical results}
\label{sec:num}

It is instructive to consider the following simple examples where we
can explicitly see the problem of equilibration of bound states and
its resolution by the introduction of auxiliary reservoirs.

\subsection{Wire with a single site}

We will first consider a system where the wire ($W$) consists of a single site
with an on-site potential and connected to one reservoir ($R$) and one
auxiliary reservoir (${\bar{R}}$). The reservoirs are one-dimensional
semi-infinite electronic lattices. The full Hamiltonian is thus given by
\bea
\CH&=& \CH_W + \CH_R +\CH_{\bar R} + {\mathcal{V}}_{R} +{\mathcal{V}}_{\bar R}
\nn ~, \\ 
\CH_W&=&V c^\dg_0 c_0 ~, \nn \\
\CH_R&=& - \g~\sum_{\a=-\infty}^{-2}~(c_\a^\dg c_{\a+1}+ c_{\a+1}^\dg c_\a ) ~,
\nn \\
{\mathcal{V}}_{R}&=& - \g'~(c_{-1}^\dg c_0 +c^\dg_0 c_{-1}) ~, \nn \\
\CH_{\bar R}&=&- \bg \sum_{\bar{\a}=-\infty}^{-2}(c_{\bar{\a}}^\dg
c_{\bar{\a}+1}+ c_{\bar{\a}+1}^\dg c_{\bar{\a}} ) ~, \nn \\
{\mathcal{V}}_{\bar R}&=& -\bg' (c_{-\bar{1}}^\dg c_0 +c^\dg_0 c_{-\bar{1}})~.
\eea
Let us first give the exact equilibrium results for the system of wire
and the single reservoir. The reservoir will be assumed to have a chemical
potential $\mu$ and temperature $T$. From App.~\ref{appeq}, we get for the
density at the single site on the wire
\bea
n^{eq}&=&n_r^{eq}+n_b^{eq} ~, \nn \\
\rm{where} \quad n_r^{eq}&=& \int_{-2}^{2} d \o ~{\g'}^2~|G_{00}^+(\o)|^2~ 
\r(\o)~ f(\o,\mu,T) ~, \nn \\
n_b^{eq}&=& \int_{-\infty}^\infty d\o~ \f{\e}{\pi}~ |G^+_{00}(\o)|^2~
f(\o,\mu,T) ~, \\
{\rm and} \quad G^+_{00}(\o) &=& \f{1}{\o+i \e -V - {\g'}^2~g^+(\o)} ~.
\label{spleq}
\eea
We need $g^+(\o)$ and $\r(\o)$ which are the reservoir Greens function and the
reservoir density of states respectively, both evaluated at the site $\a=-1$. 
The eigenvalues and eigenfunctions of the reservoir Hamiltonian are given by
\beq
\l_q ~=~ -2~ \g~ \cos q ~, \quad {\rm and} \quad \psi_q (\a) ~=~ \sqrt{2} 
\sin (q~\a) ~,
\label{epsi}
\eeq
where $q$ lies in the range $[0, \pi ]$, and $n = -1,-2, \cdots$. The wave
functions are normalized so that $\la \psi_q | \psi_{q'} \ra = \pi \d
(q-q')$. Hence we get for the required reservoir Green's function,
\bea
g^+(\o)&=& \f{1}{\g}~[~\f{\o}{2}-i~(1-\f{\o^2}{4\g^2})^{1/2}~] \quad {\rm for}
\quad |\o| < 2 \g
\nn \\
&=& \f{1}{\g}~[~\f{\o}{2 }+(\f{\o^2}{4\g^2}-1)^{1/2}~] \quad {\rm for} \quad
\o < -2 \g\nn \\
&=&\f{1}{\g}~[~\f{\o}{2 }-(\f{\o^2}{4\g^2}-1)^{1/2}~] \quad {\rm for}
\quad \o > 2 \g ~, \nn \\
\r(\o)&=&\f{1}{\g \pi}(1-\f{\o^2}{4\g^2})^{1/2} \quad {\rm for} \quad
|\o| < 2 \g \nn \\
&=& 0 \quad {\rm for} \quad |\o| > 2\g ~.
\eea
Similarly, for the auxiliary reservoir we get,
\bea
\bar{g}^+(\o)&=&\f{1}{\bg} [\f{\o}{2 \bg}-i(1-\f{\o^2}{4\bg^2})^{1/2}] \quad
{\rm for} \quad |\o| < 2 \bg \nn \\
&=&\f{1}{\bg} [\f{\o}{2 \bg}+(\f{\o^2}{4\bg^2}-1)^{1/2}] \quad {\rm for} 
\quad \o < -2 \bg \nn \\
&=&\f{1}{\bg} [\f{\o}{2 \bg}-(\f{\o^2}{4\bg^2}-1)^{1/2}] \quad {\rm for} 
\quad \o > 2 \bg ~, \nn \\
\bar{\r}(\o)&=&\f{1}{\bg \pi}(1-\f{\o^2}{4\bg^2})^{1/2} \quad {\rm for} 
\quad |\o| < 2 \bg \nn \\
&=& 0 \quad {\rm for} \quad |\o| > 2 \bg ~.
\eea

{\bf {Bound states:}} We choose the parameter values $\g=\g'=1$. For
the coupled system of wire and reservoir, we again get a continuum of states
identical to the original reservoir levels. In addition, for $V < -1$, we get
a bound state whose energy is given by $\l_b=V+1/V$, and the wave function at
the $0^{th}$ site is $\Psi_b(0)= (1-1/V^2)^{1/2}$.

\begin{figure}[htb]
\vspace{1cm}
\includegraphics[width=8.5cm]{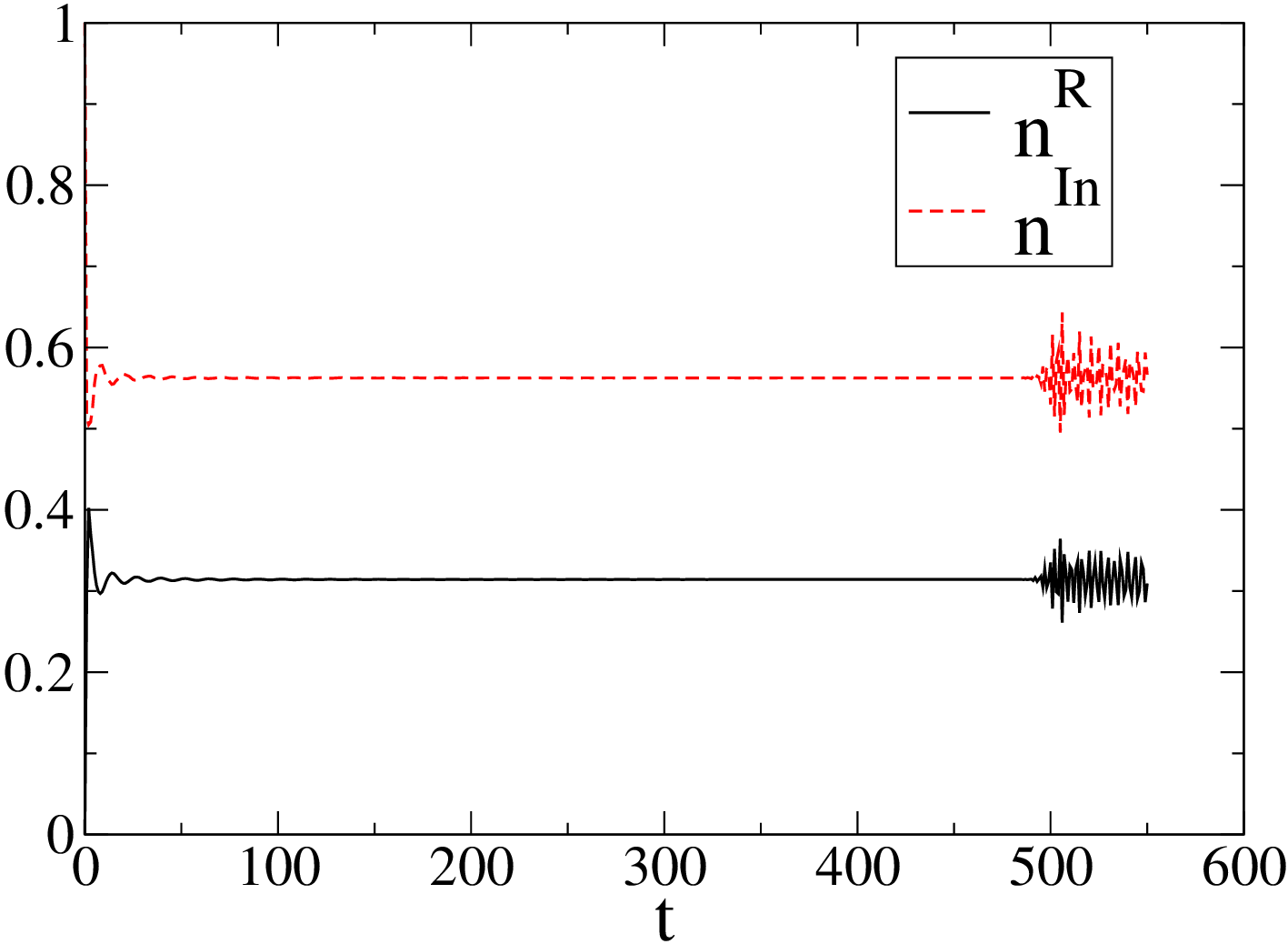}
\caption{Time evolution of the contributions of a single reservoir ($n^R$)
and the initial condition on the wire ($n^{In}$) to the density at that site,
for $\g = \g' = 1$, $\beta =2$, $\mu =0$, and $V = -2$.}
\label{figsite1}
\end{figure}

{\bf {System without auxiliary reservoir:}}
In our numerical example we take $V=-2,~\mu=0$, and $\beta=2$. Thus, from
Eq.~(\ref{spleq}), the expected equilibrium results are
\bea
n^{eq}_r &=& \int_{-2}^{2} d\o~ \f{\r(\o) f(\o,\mu,T)}{|\o-V-g^+(\o)|^2} ~=~ 
0.1641... ~, \nn \\
n^{eq}_b &=& \Psi_b(0)^2 f(\l_b,\mu,T) ~=~ 0.7449... ~, \nn \\
n^{eq} &=& n^{eq}_r+n^{eq}_b ~=~ 0.9091... ~.
\label{spleq2}
\eea
We have numerically solved the equations of motion for a reservoir with
$L=500$ sites. In Fig.~\ref{figsite1}, we show the contribution of the 
reservoir to the density at the $0^{\rm{th}}$ site
as a function of time by a solid line. We find that it quickly
settles at a value of about $n^R=0.314$. The contribution coming from an
initial density equal to one at site $0$, namely, $\lim_{t \to \infty}
G^+_{00} (t) G^-_{00} (t)$ in Eq.~(\ref{nlm}), is shown by a dashed line.
This does not vanish with time but settles at a value of about $n^{In}=0.563$.
This means that the steady state density on the wire depends on the initial 
density. Using the results of Sec.~\ref{sec:gensol} we can
understand the different contributions to the density. From Eq.~(\ref{Glt}),
we get the contribution from the initial density as 
\bea
n^{In} ~=~ |\Psi_b(0)|^4 ~=~ 0.5625~,
\eea
which is close to the numerically obtained result. From Eq.~(\ref{rescont}), 
we see that there are two parts to the contribution from the reservoir levels.
The first part is a contribution to the density at site $0$ arising from
the extended states; this is identical to the equilibrium result
$n_{ext}^{R}=n^{eq}_r=0.1641...$. The second part is from the
bound state; from Eq.~(\ref{rescont}) this is given by
\bea
n_{bnd}^{R} ~=~ |\Psi_b(0)|^4 ~\int_{-2}^{2} d\o~
\f{\r(\o)}{(\o-\l_b)^2}~f(\o,\mu,T) ~=~ 0.1500... ~. \nn 
\eea
Hence we get $n^R=n_{ext}^{R}+n_{bnd}^{R}=0.3141...$.
We note in Fig.~\ref{figsite1} that for large times which are of the order of
the reservoir size, both the contributions start deviating from their steady
values; we will comment more on this below.

\begin{figure}[htb]
\vspace{1cm}
\includegraphics[width=8.5cm]{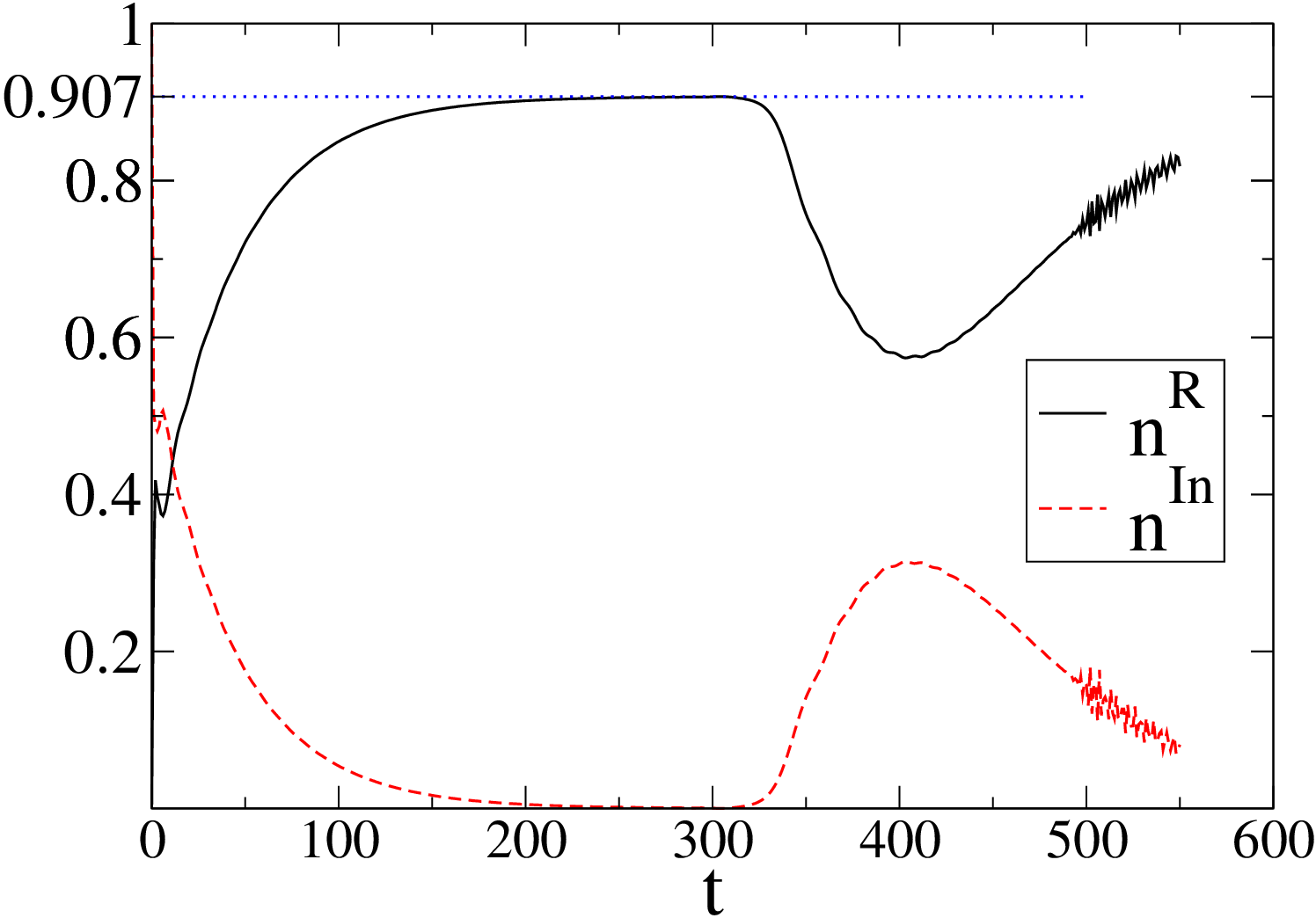}
\caption{Time evolution of the contributions of the two reservoirs ($n^R$)
and the initial conditions on wire ($n^{In}$) to the density at that site,
for $\g = \g' = 1$, ${\bar \g} = 2$, ${\bar \g'} = 0.2$, $\beta =2$,
$\mu = 0$, and $V = -2$.}
\label{figsite12r}
\end{figure}

{\bf {System with auxiliary reservoir:}}
We now introduce an auxiliary reservoir with the same temperature, chemical
potential and number of sites as the original reservoir. However, we take the
hopping amplitude in this reservoir to be ${\bar \g} = 2$, so that its
bandwidth of $[-4 , 4]$ includes the energy of the bound state of the original
system. We take the coupling of the auxiliary reservoir to site 0 to be
${\bar \g'} = 0.2$. We now solve the equations of motion of this new system
consisting of two reservoirs and one site. Fig.~\ref{figsite12r}
shows the total contribution of the two reservoirs by a solid line, and
the contribution coming from an initial density of one at site 1 by a dashed
line. We see that the solid line approaches a value of $0.907$ (this deviates
slightly from the equilibrium value of $0.909$ because $\bar \g'$, though
small, is not zero), while the dashed line vanishes around the same time.
Using the results in Sec.~\ref{sec:reso}, we can compute the contributions of
the reservoir and the auxiliary reservoir to the net density. We get
\bea
n^R ~=~ \int_{-2}^2 d \o~ \f{ \r (\o) f(\o,\mu,T)}{|\o-V-g^+(\o)-\bg'^2 
\bar{g}^+(\o)|^2} ~=~ 0.1614... ~, \nn \\
n^{\bar R} ~=~ \int_{-4}^4 d \o~ \f{ \bg'^2 \bar{\r} (\o) f(\o,\mu,T)}{|\o-V-
g^+(\o)-\bg'^2 \bar{g}^+(\o)|^2} ~=~ 0.7458... ~.
\eea
Hence the total density is $n=n^R+n^{\bar R}=0.9072...$ which is
consistent with the value obtained from the numerics. Comparing with
Eq.~(\ref{spleq2}) we note that the auxiliary reservoir only contributes to
equilibration of the bound state.

\begin{figure}[htb]
\vspace{1cm}
\includegraphics[width=8.5cm]{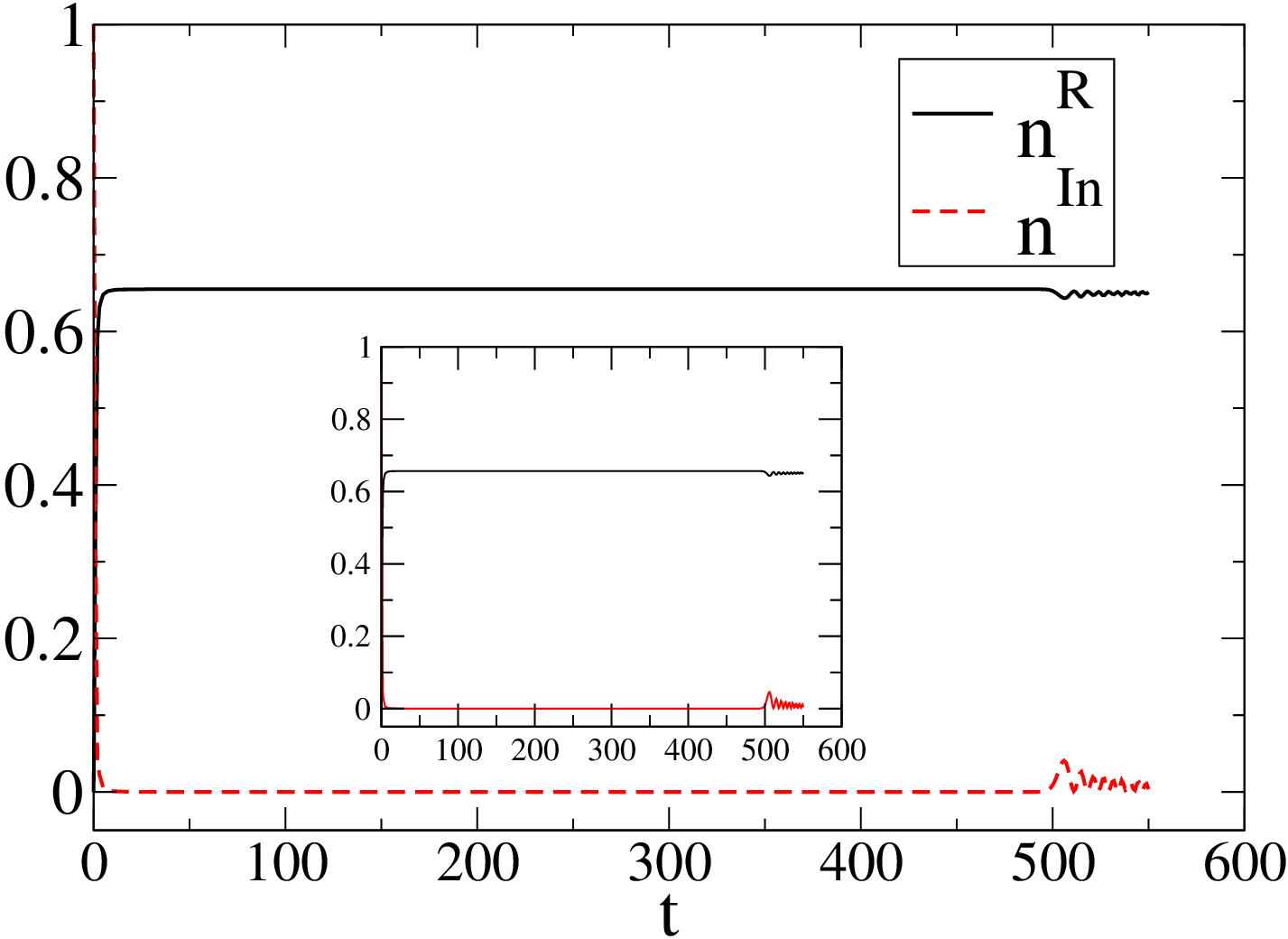}
\caption{Time evolution of the contributions of a single reservoir ($n^R$)
and the initial conditions on wire ($n^{In}$) to the density at that site,
for $\g = \g' = 1$, $\beta =2$, $\mu = 0$, and $V = -0.5$. The inset shows
the time evolution for the case with two reservoirs.}
\label{nobndfig}
\end{figure}

In Fig.~\ref{nobndfig}, we show an example where the potential $V=-0.5$ is
such that there are no bound states. In this case we find, as expected, that
the wire equilibrates even with a single reservoir. Adding an auxiliary
reservoir leaves the density essentially unchanged.

The time scale of approaching these steady values is of the order of $\tau = 
{\bar \g}/ {\bar \g}^{'2} = 50$. This is due to the fact that the self-energy
of the auxiliary reservoir is proportional to ${\bar \g}^{'2} /{\bar \g}$. This
governs the rate at which the auxiliary reservoir fills up site 1, and it is
inversely proportional to $\tau$. For times $t >> \tau$, we obtain the correct
equilibrium value of the density at site 0 (up to a small error due to
the finiteness of $\bar \g'$). Further, the vanishing of the dashed line 
means that the steady state density at site 0 does not depend on the initial
density at that site. Once again, we see in Fig.~\ref{figsite12r} that for 
large times of the order of the reservoir sizes, both the 
contributions start deviating from their steady values.

The deviations which begin appearing in Figs.~\ref{figsite1}-\ref{nobndfig} at
large times of the order of the reservoir sizes can be understood as follows. 
After the wire is suddenly connected to one end of a reservoir at time $t=0$,
there is a recurrence time of the reservoir which is given by the time it
takes for the effect of that connection to propagate to the other end of the
reservoir, and then return to the wire. If the reservoir has a size $L$,
and the Fermi velocity in the reservoir is $v_F$ (this is equal to
$2 \g \sin k_F$ and $2 {\bar \g} \sin k_F$ for the original and auxiliary
reservoirs respectively), the recurrence time is given by $2L/v_F$.
Since we have taken $\mu = 0$ (half-filling), we have $k_F = \pi /2$, and
$v_F = L/\g$ and $L/{\bar \g}$ in the two reservoirs respectively. In
Figs.~\ref{figsite1}-\ref{nobndfig}, we see deviations occurring at a time 
given by $L=500$ due to the original reservoir, while in Fig.~\ref{figsite12r},
we also see deviations occurring at a time equal to $L /{\bar \g} = 250$ due 
to the auxiliary reservoir.

The presence of the recurrence times imply that one must take the limits of
certain quantities going to infinity in a particular order, in order to
numerically obtain the correct steady state values of the density matrix.
The coupling $\bar \g'$ of the auxiliary reservoirs to the wire must be
taken to zero so that the steady state values do not
depend on the details of those reservoirs (at least in the equilibrium case);
this means that $\tau$ must go to infinity. On the other hand,
the recurrence time $2L/v_F$ must also go to infinity. The steady
state will then exist for times $t$ which satisfy $\tau << t << 2L/v_F$.

{\bf {Approach to equilibrium:}} Let us briefly discuss the way in which
equilibrium is approached at times which are small compared to the recurrence
times discussed above. One again, there are two different time scales here,
one for the original reservoirs and the other for the auxiliary reservoirs
(if there is a bound state present). For the original
reservoirs, let us consider an integral of the form given in the
second term in Eq.~(\ref{large}), namely, $\int_{E_1}^{E_2} d\o ~\rho 
(\o) e^{-i \o t}$, where the function $\rho (\o)$ has no singularities
in the range $[E_1 ,E_2]$. Let us also suppose that $\rho (\o)$ has
the power-law forms $(\o - E_1)^{\a_1}$ and $(E_2 - \o)^{\a_2}$ at the
two ends of the integral, where $\a_1 ,~\a_2 > -1$ so that the integral
exists. Then one can use the method of steepest descent to show that in the
limit $t \to \infty$, the above integral only gets a contribution from the
end points, and those contributions vanish as $e^{-iE_1 t} /t^{\a_1 + 1}$
and $e^{-iE_2 t} /t^{\a_2 + 1}$ respectively \cite{bender}. For the original
reservoirs with $\g = \g' =1$, this approach to equilibrium occurs at times
of the order $1/\g$, and it is therefore hard to see the power-law fall-off
in Figs.~\ref{figsite1} and \ref{fig2site}. For 
the auxiliary reservoirs, the time scale of equilibration is given by a
different expression if there is a bound state present. The relevant integral
is then given by $\int_{{\bar E}_1}^{{\bar E}_2} d\o ~\rho (\o) e^{-i \o t}$,
where the function $\rho (\o)$ is large in the vicinity of the bound state
energy $\l_b$, namely, $\rho (\o) \sim ({\bar \g}^{'2} /{\bar \g}) /[(\o - 
\l_b)^2 + ({\bar \g}^{'2} /{\bar \g})^2 ]$ where ${\bar \g} /{\bar \g}^{'2}$ is
large. We then see that the dominant contribution to the integral comes from 
the vicinity of $\o = \l_b$ and is given by $e^{-t~ {\bar \g}^{'2} /
{\bar \g}}$. We can see this exponential decay in Fig.~\ref{figsite12r}, 
till the effect of the shortest recurrence times starts becoming visible. 

\subsection{Wire with two sites}

\begin{figure}[htb]
\includegraphics[width=8.5cm]{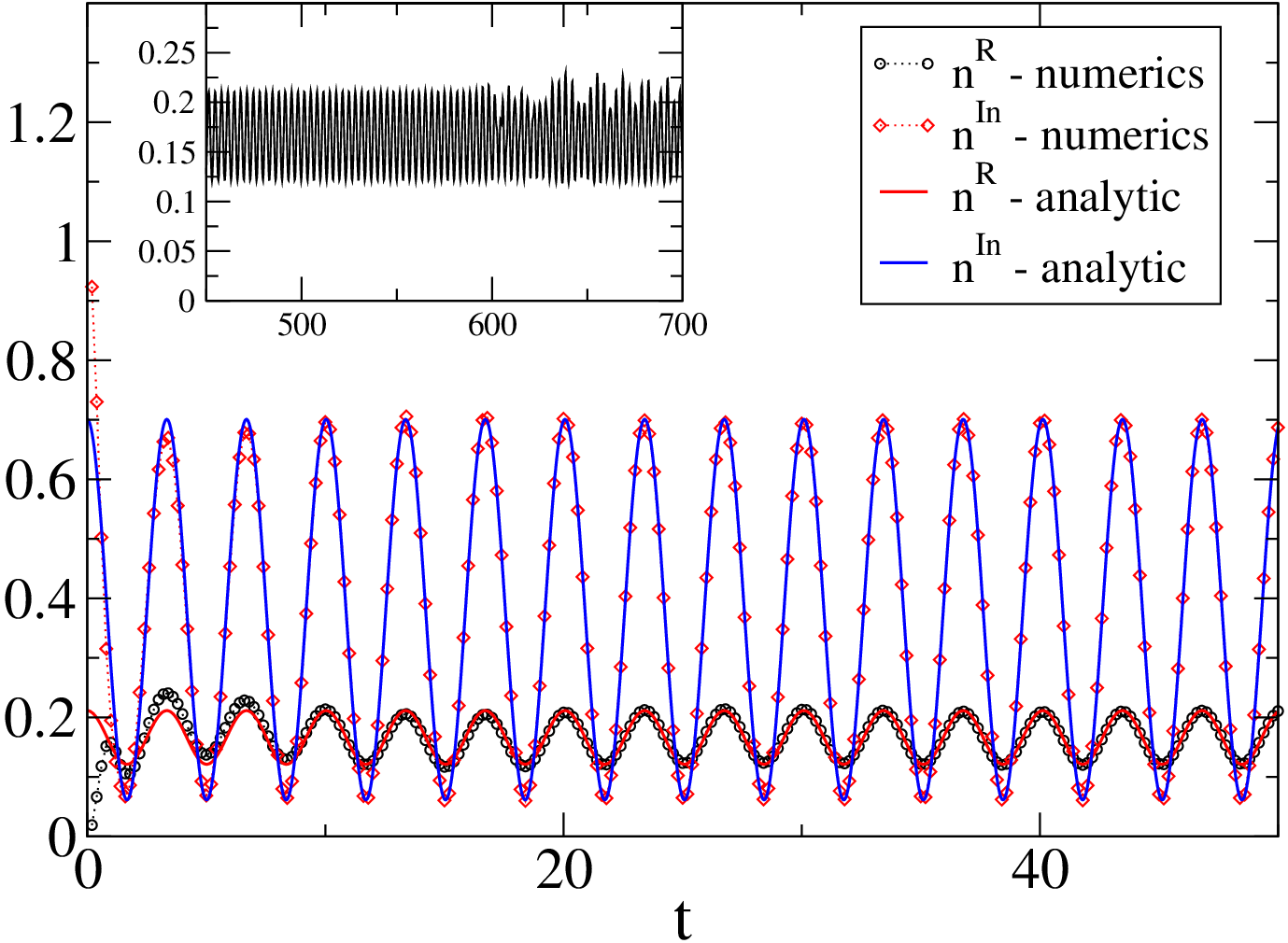}
\caption{Time evolution of the contributions of the reservoir ($n^R$) 
and initial conditions on wire ($n^{In}$) to the density at the first
site of a two-site system, for $\g=\g'=1$, $\beta=2,~\mu=0$ and $v=-3$. The 
analytic predictions are also plotted.}
\label{fig2site}
\end{figure}

\begin{figure}[htb]
\vspace{1cm}
\includegraphics[width=8.5cm]{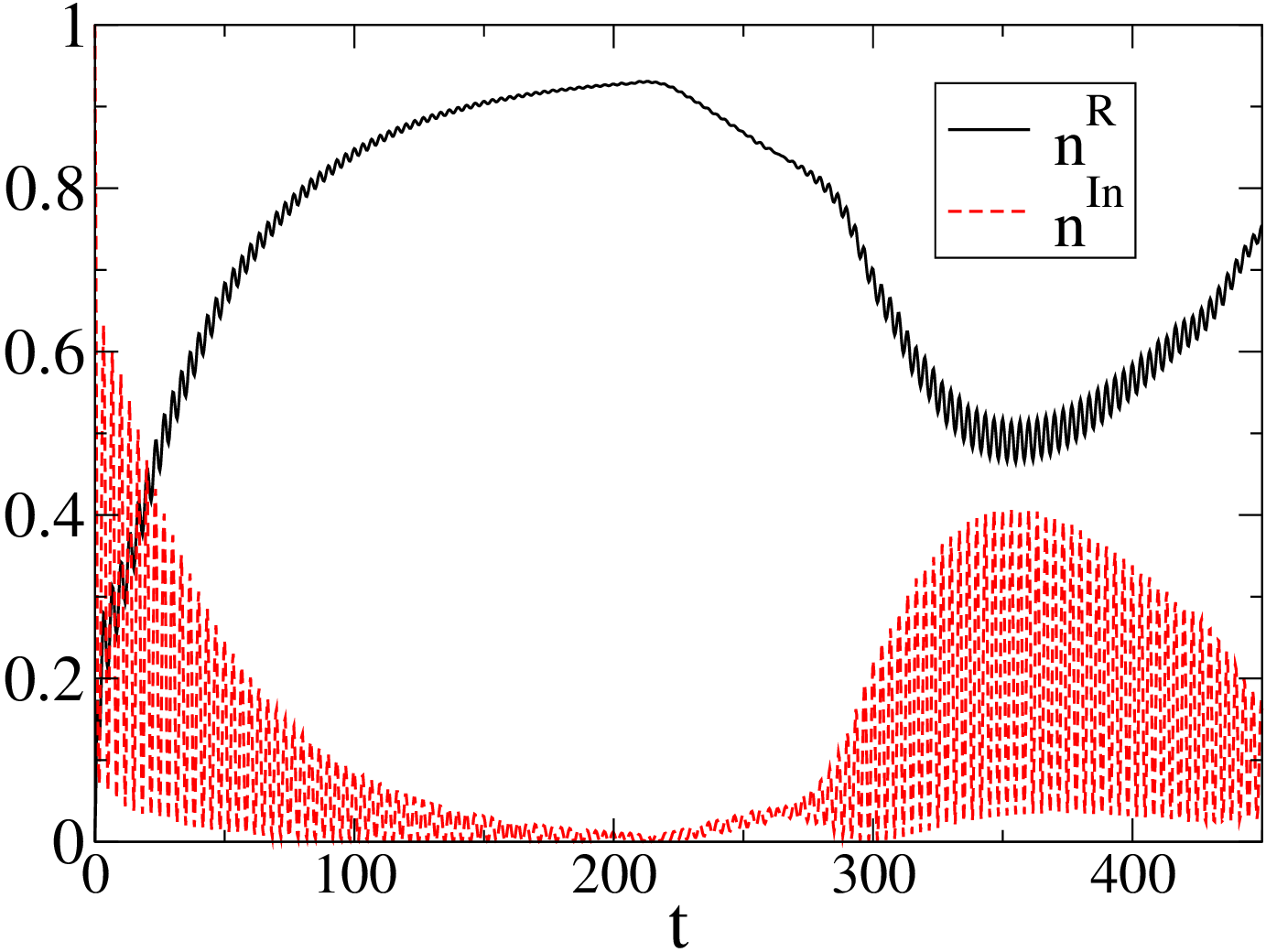}
\caption{Time evolution of the contributions of the two reservoirs ($n^R$) 
and the initial density on site ($n^{In}$) to the density at the first site 
of a two-site system, for $\g = \g' = 1$, ${\bar \g} = 3$, ${\bar \g'} = 0.3$,
$\beta =2$, $\mu = 0$, and $V = -3$.}
\label{fig2site2R}
\end{figure}

In the presence of two bound states, the steady state properties can be time 
dependent. We now illustrate this with the example of a two-site wire. 
We take both sites (labeled 0 and 1) to have an on-site potential $V=-3$. The
site 0 is connected to a one-dimensional semi-infinite reservoir (going
from $-1$ to $-L$ as in the previous example), while site 1 is only 
coupled to site 0. The Green's function is thus
\bea
G^+(\o) = \left( \begin{array}{cc}
\o-V-g^+(\o) & 1 \cr
 1 & \o-V \end{array} \right)^{-1}
\eea
Let us look at the density at site $0$ in the long time limit. From the 
equation of motion solution in Sec.~\ref{sec:gensol}, we expect the density to
have contributions from both the initial density matrix on the wire ($n^{In}$)
and from the reservoir ($n^R$). We choose the initial wire density matrix to be
diagonal, with $\la c^\dg_0 c_0 \ra =1$. Then from Eq.~(\ref{large}) we get
\bea
n^{In} ~=~ |\Psi_{b_1}(0)|^4 ~+~ |\Psi_{b_2}(0)|^4 ~+~ 2 ~|\Psi_{b_1}(0)|^2 
|\Psi_{b_2}(0)|^2 ~\cos{(\l_{b_1}-\l_{b_2})t} ~.
\label{eq2s1} 
\eea
The contribution from the reservoirs consists of two parts: one
corresponding to the extended states $n^{R}_{ext}$, and the other to
the bound states $n^{R}_{bnd}$. These are given by
\bea
n^{R}_{ext} &=& \int_{-2}^{2} d\o ~|G^+_{00}|^2 ~\r(\o) ~f(\o,\mu,T) \nn \\
n^{R}_{bnd} &=& |\Psi_{b_1}(0)|^4 ~\int_{-2}^{2} d\o ~\f{\r(\o)
f(\o,\mu,T)}{(\o-\l_{b_1})^2} ~+~ |\Psi_{b_2}(0)|^4 ~\int_{-2}^2 d\o~\f{\r(\o)
f(\o,\mu,T)}{(\o-\l_{b_2})^2} ~, \nn \\
& & + ~2~ |\Psi_{b_1}(0)|^2 |\Psi_{b_2}(0)|^2 ~\cos{(\l_{b_1}-\l_{b_2})t}~
\int_{-2}^2 d\o ~\f{ \r(\o) f(\o,\mu,T)}{(\o-\l_{b_1})(\o-\l_{b_2})} ~.
\label{eq2s2}
\eea
The bound state eigenvalues are obtained by solving, for $\o=-2 \cosh{\a} <0 $,
the equation
\bea
Det ~[ \left( \begin{array}{cc}
\o-v-g^+(\o) & 1 \cr
1 & \o-v \end{array} \right) ] ~=~ 0 ~.
\eea
With $g^+(\o)=-e^{-\a}$ this gives the equation
\bea
e^{3 \a}+ 2 v e^{2 \a} + v^2 e^{\a} + v ~=~ 0 ~, \nn
\eea
which, for $V=-3$, has two solutions with $\a > 0$. These give the eigenvalues
$\l_{b_1}=-2.257...$ and $\l_{b_2}=-4.137...$. The eigenvectors can also be 
found easily after normalizing them over the entire system. At site 0, we get
\bea
|\Psi_b(0)|^2 ~=~ \f{(1-e^{-2\a}) (v+2 \cosh{\a})^2}{(1-e^{-2\a})+ (v+2 
\cosh{\a})^2} ~. \nn
\eea
This gives $|\Psi_{b_1}|^2=0.2947...$ and $|\Psi_{b_2}|^2 =0.5421...$.
Then from Eq.~(\ref{eq2s1}) and Eq.~(\ref{eq2s2}) we get
\bea
n^{In}&=& 0.381 ~+~ 0.3196 ~\cos (1.879 t) ~, \nn \\
n^{R}_{ext}&=& 0.1097 ~, \nn \\
n^{R}_{bnd}&=& 0.05654 ~+~ 0.045 ~\cos (1.879 t) ~.
\eea
We verify these results by an exact numerical solution of the time evolution
of the system with the two-site wire and a one-dimensional reservoir with
$L=600$ sites. In Fig.~\ref{fig2site}, we compare the analytic results with
the numerical solution and find very good agreement between the two. The
inset shows the long time behavior; we see that the effect of the finite
size of the reservoir shows up at the recurrence time $2L/v_F = 600$.
In Fig.~\ref{fig2site2R}, we show the effect of adding
a weakly coupled ($\bg'=0.3$) auxiliary reservoir with $\bg=3$. As expected, 
in this case there is no contribution from the initial density of the wire,
and the reservoir contribution gives the equilibrium value (till the effect of
the recurrence time of the auxiliary reservoir, $L/{\bar \g} = 200$, shows up).

\section{Interacting systems}
\label{sec:inter}

Let us briefly consider how interactions can be studied within the NEGF
formalism. We note that
this formalism works with a one-particle Hamiltonian, e.g., the self-energy
in Eq.~(\ref{self}) is given in terms of the energy of a single electron
which is entering or leaving the wire. One way to deal with interactions is
therefore to do a Hartree-Fock (HF) decomposition. For instance, if we have
a Hubbard model with an on-site interaction between spin-up and spin-down
electrons of the form $U c_{n\up}^\dg c_{n\up} c_{n\dn}^\dg c_{n\dn}$,
we can approximate it by as
\beq
U ~[~\la c_{n\up}^\dg c_{n\up} \ra ~c_{n\dn}^\dg c_{n\dn} ~+~
c_{n\up}^\dg c_{n\up} ~\la c_{n\dn}^\dg c_{n\dn} \ra ~-~ \la c_{n\up}^\dg
c_{n\up} \ra ~\la c_{n\dn}^\dg c_{n\dn} \ra ~]~.
\eeq
We see that the first two terms modify the on-site potential. A 
self-consistent NEGF calculation can then be implemented as follows 
\cite{agarwal}.

\begin{itemize} 
\item{Start with the Hamiltonian with no interactions, and calculate the
density matrix. The diagonal elements of the density matrix give the
densities at different sites.}
\item{Use the HF approximation to compute the Hamiltonian with interactions,
and use that to calculate the density matrix again.}
\item{Repeat the previous step till the density matrix stops changing.}
\item{Use the converged density matrix to calculate the site densities or the
current.}
\end{itemize}

The important point is that in an interacting system, the density affects the
on-site potential and therefore the current. If there is a bound state
present, it will affect the current by modifying the density, even though the
bound state does not directly contribute to the current. Hence a proper 
treatment of bound states is necessary in order to calculate the current.
Numerically, it is found that the presence of a bound state has a significant 
effect on the conductance of an interacting system \cite{agarwal}.

The effect of on-site Coulomb interactions on the conductance through a 
quantum dot has been studied earlier in Ref.~\onlinecite{meir2} using a 
self-consistent truncation of the equations of motion for the Green's 
function. The density on the dot plays an important role in that analysis; 
it is therefore clear that the presence of a bound state would affect the 
conductance.

\section{Discussion}
\label{sec:disc}

In conclusion, we have pointed out that in the presence of bound states and
no additional mechanisms for equilibration, the NEGF formalism gives a 
non-unique density matrix. We have shown that the equation of motion approach,
which gives a simple and straightforward method of deriving the results of 
NEGF for non-interacting systems, also provides a clear understanding of the 
ambiguity in the density matrix when bound states are present.

We have then presented a way of resolving the ambiguity which arises when 
there are bound states. Namely, for each reservoir, we introduce an
auxiliary reservoir with the same temperature and chemical potential;
the auxiliary reservoir is taken to have a larger bandwidth (so that the
bound state energy lies within that bandwidth), and a much smaller
coupling to the wire than the original reservoir.
We then find that the contribution of the bound state is completely 
determined by the properties of the auxiliary reservoirs. 
If we let the couplings of the auxiliary reservoirs tend to zero,
we get the correct equilibrium density matrix including the
contribution from the bound states. In particular, we find that
the bound state contribution is unique and independent of the details of 
the auxiliary reservoirs in the equilibrium case. However we again
find that in the non-equilibrium case, the bound state contribution 
is not unique and depends on the details of the auxiliary reservoirs 
and the way in which the limit of zero coupling is taken. 

One might suspect that the sudden switching on of the wire-reservoir
interactions could be a reason for the problem of equilibration. However we 
have also studied the case where the coupling is switched on adiabatically 
and verified that the equilibration problem remains.

For non-interacting systems, bound states do not directly contribute to the 
current, and so transport properties are not affected. However in the presence
of electron-electron interactions, a simple mean-field treatment suggests that
bound states affect the local density which in turn affects the current. Thus
transport properties of interacting systems can be affected in a 
non-trivial way in the presence of bound states. 

\vskip .5 true cm
\centerline{\bf Acknowledgments}
\vskip .5 true cm

DS thanks Supriyo Datta and Amit Agarwal for stimulating discussions, and the 
Department of Science and Technology, India for financial support under 
projects SR/FST/PSI-022/2000 and SP/S2/M-11/2000. AD thanks Subhashish
Banerjee for a critical reading of the manuscript and also for bringing to 
our notice some relevant references.

\appendix

\section{Green's function properties}
\label{appG}

The single particle Green's function for the coupled system of wire and 
reservoirs is given by
\bea
\CG^+(t)= -i e^{-i H t} \theta(t) ~.
\eea
It satisfies the equation of motion
\bea
i\f{\p \CG^+}{\p t}-H \CG^+= \d(t) ~.
\eea
The Fourier transform $\CG^+(\o)= \int_{-\infty}^\infty dt~
\CG^+(t)e^{i \o t}$ is thus given by
\bea
\CG^+(\o)=\f{1}{(\o+i \e) \hat{I} - \hat{H}} ~, \label{fullg}
\eea
which can also be represented as
\bea
\CG^+_{rs}(\o) &=&\sum_Q \f{\Psi_Q(r) \Psi_Q^*(s)}{\o+i \e-\l_Q} \nn \\
&=& \sum_Q \f{\Psi_Q(r) \Psi_Q^*(s)}{\o-\l_Q} ~-~ i \pi ~\sum_Q ~\Psi_Q(r)
\Psi_Q^*(s) ~\d(\o-\l_Q) ~, \nn
\eea
where $\Psi_Q$ and $\l_Q$ respectively denote eigenfunctions and eigenvalues
of the full system. Now what we need is the part of the full system
Green's function defined between points on the wire. Let us try to
express this part, which we will genote by $G^+(\o)$, in terms of
the isolated reservoir Green's functions given by:
\bea
g^+_L(\o)=\f{1}{\o+i \e-H^L} ~,~~~~ g^+_R(\o)=\f{1}{\o+i \e -H^R}
\eea
of the isolated reservoirs. We rewrite the Green's function equation, 
Eq.~(\ref{fullg}), by breaking it up into various parts corresponding to the 
wire and reservoirs. Thus we get
\bea
\left( \begin{array}{ccc}
(\o+i \e) \hat{I}- H^W & -~ V^L & -~ V^R \cr
-~ V^{L^\dg} & (\o+i \e) \hat{I}-H^L & 0 \cr
-~ V^{R^\dg} & 0 & (\o+i \e) \hat{I}-H^R \cr
\end{array} \right) \left( \begin{array}{ccc}
G^+ & G^+_{WL} & G^+_{WR} \cr
G^+_{LW} & G^+_L & G^+_{LR} \cr
G^+_{RW} & G^+_{RL} & G^+_R \cr
\end{array} \right) = \left( \begin{array}{ccc}
\hat{I} & 0 & 0 \cr
0 & \hat{I} & 0 \cr
0 & 0 & \hat{I} \cr
\end{array} \right). \nn
\eea
This gives the following equations:
\bea
[(\o+i \e) \hat{I}- H^W] G^+ ~-~ V^L G^+_{LW} ~-~ V^R G^+_{RW} &=& \hat{I} ~,
\nn \\
-~ V^{L^\dg}~ G^+(\o) ~+~ [~(\o +i \e) \hat{I}-H^L~]~ G^+_{LW} &=& 0 ~, \nn \\
-~ V^{R^\dg}~ G^+(\o) ~+~ [~(\o +i \e) \hat{I}-H^R~]~ G^+_{RW} &=& 0 ~.
\eea
Solving the last two equations gives $G^+_{LW}(\o)= g^+_L(\o)~ V^{L^\dg}~ 
G^+(\o)$ and $G^+_{RW}(\o)= g^+_R(\o)~ V^{R^\dg}~ G^+(\o)$. Using this in 
the first equation then gives
\bea
&& [~(\o+i \e) \hat{I}- H^W- V^L g^+_L(\o) ~V^{L^\dg} - V^R~ g_R^+(\o) ~
V^{R^\dg}~]~ G^+(\o) ~=~ \hat{I} ~, \nn \\
&& \Rightarrow ~G^+(\o) ~=~ \f{1}{\o+i \e -H^W - \S^+_L-\S^+_R } ~, 
\label{Egrw} \\
&& {\rm where}~~~\S^+_L(\o) ~=~ V^L g^+_L(\o) ~V^{L^\dg} ~,~~~~\S^+_R(\o) ~=~
V^R~ g_R^+(\o) ~V^{R^\dg} \nn~. 
\eea

\section{Equilibrium properties}
\label{appeq}

Let us calculate the expectation value of the density matrix for the
case where the entire system of wires and reservoirs are described by
a grand canonical ensemble at chemical potential $\mu$ and temperature
$T$. For points on the wire this is given by
\bea
n^{eq}_{lm} ~=~ \la~ c^\dg_m~ c_l~ \ra_{eq}
&=& \sum_Q \Psi_l(Q) \Psi_m^*(Q) ~f(\l_Q, \mu,T) \label{neqsumq} \\ 
&=& \int_{-\infty}^\infty d \o ~\sum_Q ~\Psi_l(Q) \Psi_m^*(Q) ~\d( \o
-\l_Q) ~f(\o,\mu,T) \nn \\
&=& \int_{-\infty}^\infty d \o ~\f{1}{2 \pi i} [G^-(\o)-G^+(\o)]_{lm} ~
f(\o, \mu,T) ~.
\eea
Now from Eq.~(\ref{Egrw}) we get
\bea
(G^+)^{-1}- (G^-)^{-1} &=& (\S^-_L - \S^+_L) + (\S^-_R-\S^+_R)+2 i \e ~=~ 
2 \pi i (\G_L +\G_R) +2i \e ~, \nn \\
\Rightarrow ~G^- - G^+ &=& 2 \pi i~ G^+ ( \Gamma_L + \Gamma_R )~ G^- ~+~ 
2i \e~ G^+ G^- ~.
\eea
Therefore we get for the equilibrium density matrix
\bea
n^{eq}_{lm}= \int_{-\infty}^\infty d \o~ [G^+ ( \Gamma_L + \Gamma_R ) 
G^-]_{lm} ~f(\o,\mu,T) + \int_{-\infty}^\infty d \o~ \f{\e}{\pi}
~[G^+ G^-]_{lm}~ f(\o, \mu,T) ~.
\label{Eeqdns}
\eea
The second part is non-vanishing only if the equation
\bea
Det ~[\o \hat{I}- H^W-\S^+_L(\o)-\S^+_R(\o)]~=~ 0
\eea
has solutions for real $\o$. These correspond to the {\emph {bound states}} of
the coupled system. As usual, the limit $\e \rightarrow 0$ is implied in 
Eq.~(\ref{Eeqdns}). The second term in Eq. (\ref{Eeqdns}) survives in that 
limit if there is a bound state, since $\e~ G^+ G^- \sim \e /[(\o - \l_b)^2 +
\e^2] = \pi \d (\o - \l_b)$ in that case. In fact it is easy to see that the
second term in Eq.~(\ref{Eeqdns}) reduces precisely to the following form:
\bea
 \int_{-\infty}^\infty d \o~ \f{\e}{\pi} ~[G^+ G^-]_{lm}~ f(\o, \mu,T) ~=~ 
\sum_b ~\Psi_l(b) \Psi_m^*(b) ~f(\l_b,\mu,T)~,
\eea
which can also be seen directly from Eq.~(\ref{neqsumq}).

\end{document}